\documentclass[runningheads]{llncs}
\usepackage[T1]{fontenc}
\usepackage[margin=2.6cm]{geometry} 
\usepackage[section]{placeins}
\usepackage{color, array}
\usepackage{authblk}
\usepackage{placeins}
\usepackage{float}
\usepackage{wrapfig}
\usepackage{array} 
\usepackage{subcaption}
\usepackage{comment}
\usepackage{graphicx}
\usepackage{siunitx}
\usepackage{amsmath}
\usepackage{booktabs}
\usepackage{tabularx}

\begin{document}

\title{Unsupervised Particle Tracking with Neuromorphic Computing}

%
\author{
Emanuele Coradin\inst{1,2} \and
Fabio Cufino\inst{3} \and
Muhammad Awais\inst{1,2,7} \and
Tommaso Dorigo\inst{4,1,2,5,7} \and
Enrico Lupi\inst{1,2} \and
Eleonora Porcu\inst{3} \and
Jinu Raj\inst{6}\and
Fredrik Sandin\inst{4,7}\and
Mia Tosi\inst{1,2,7}
}

%
\institute{
INFN, Sezione di Padova, Via F. Marzolo 8, 35131 Padova, Italy \and
Università di Padova, Dipartimento di Fisica e Astronomia, Via F. Marzolo 8, 35131 Padova, Italy \and
Università di Bologna, Dipartimento di Fisica, via Irnerio, Bologna, Italy \and
Luleå University of Technology, 971 87 Luleå, Sweden \and
Universal Scientific Education and Research Network, Italy \and
Central University of Tamil Nadu, Thiruvarur, 610 001, India \and
MODE Collaboration, \url{https://mode-collaboration.github.io} \\
\email{emanuele.coradin01@gmail.com, fabio.cufino@studenti.unipd.it}
}
\authorrunning{Emanuele C., Fabio C. et al.}
\maketitle              

\begin{abstract}
We study the application of a neural network architecture for identifying charged particle trajectories via unsupervised learning of delays and synaptic weights using a spike-time-dependent plasticity rule. In the considered model, the neurons receive time-encoded information on the position of particle hits in a tracking detector for a particle collider, modeled according to the geometry of the Compact Muon Solenoid Phase II detector. We show how a spiking neural network is capable of successfully identifying in a completely unsupervised way the signal left by charged particles in the presence of conspicuous noise from accidental or combinatorial hits. These results open the way to applications of neuromorphic computing to particle tracking, motivating further studies into its potential for real-time, low-power particle tracking in future high-energy physics experiments.


\end{abstract}



\section{Introduction} 

The aspiration for higher spatial resolution of particle detectors, to enhance the scientific potential of High Energy Physics (HEP) experiments, leads to an extremely large data volume~\cite{HL-HLC}.
Conventional computing solutions alone struggle with the demand of online identification and reconstruction of particle signals~\cite{dorigo2023mode,mehonic2024review}. Another significant challenge lies in the temporal realm.
Given that particles close to light speed traverse \SI{3}{\centi\meter} in just 100 picoseconds, the exploitation of time patterns of detected signals requires sensitivity to sub-nanosecond time intervals. Current detectors neglect the temporal information generated by the passage of particles through sensitive material, while it could in principle help extracting further information.
The integration of fast online dimensional reduction and spatiotemporal pattern recognition using parallel analog-digital neuromorphic computing architectures at the detector end may allow us to overcome these limitations.

\subsection{Neuromorphic Computing} 


As digital computing technologies approach their physical and architectural limits, alternative mixed-signal processing methods that exploit intrinsic physical properties of materials are investigated ~\cite{jaeger2023naturecomms}, a concept sometimes referred to as 'in-materia' and 'analog in-memory' computing.
A prominent example is neuromorphic computing and engineering~\cite{mead2020natureel}, where sensors, processors and cybernetic system architectures are developed using biological brains as inspiration.
Brains process massively parallel event-based representations of uncertain information, providing new insights into how spatiotemporal detector signal patterns can be efficiently encoded and compressed.

Neuromorphic solutions imitate mixed-signal circuits and architectural motifs inspired by the brain to improve efficiency, robustness and learning~\cite{Kudithipudi2025,mehonic2024neuromorphic}.
Brains demonstrate orders-of-magnitude more energy-efficient learning and intelligence solutions and show that relatively slow, stochastic, and plastic circuits can promote robust processing of unstructured and noisy avalanches of sensor information.
For example, the FPGA-based neuromorphic supercomputer DeepSouth presently under construction can reach above 200 trillion synaptic operations per second at \SI{40}{\kilo W} power, while a human brain with a comparable number of neurosynaptic operations (and higher complexity) requires about \SI{20}{W}~\cite{deepsouthwww}.

Although the dynamical time constants of biological neurons are matched to the long, behaviorally relevant, timescales of milliseconds or more, the general pattern learning capacity of such neurosynaptic circuits can be generalized to other environments.
By implementing circuits that mimic useful aspects of the neurosynaptic mixed-signal dynamics in, {\it e.g.}, nanoscale semiconductor and photonic devices, neuromorphic circuits with sub-nanosecond time constants can be realized ~\cite{winge2020,wittenbecher2021,Winge2023}.
This opens new opportunities for developing efficient event-triggered information sampling and spiking neural network processing solutions ~\cite{Kudithipudi2025,Nilsson2023a} for high-energy physics' detector readouts.

\subsection {Spiking Neural Networks} 

Spiking Neural Networks (SNNs) are used to model biological neurons using differential equations describing neurosynaptic dynamics at some spatial and temporal approximation level. Unlike conventional artificial neural networks (ANNs), which use continuous values to represent the activation of neurons, SNNs are based on discrete spikes that neurons generate in response to incoming stimuli. These spikes occur at precise points in time, adding a temporal dimension to neural processing that enhances the model's capacity to asynchronously process time-dependent information efficiently, see for example ~\cite{Nilsson2023b}. This spike-based approach makes SNNs particularly well-suited for event-based spatiotemporal processing and neuromorphic computing, where neurosynaptic dynamics and asynchronous parallel processing are efficiently implemented using specialized circuits and materials.

\begin{figure}[!h]
    \centering
    \includegraphics[width=0.6\linewidth]{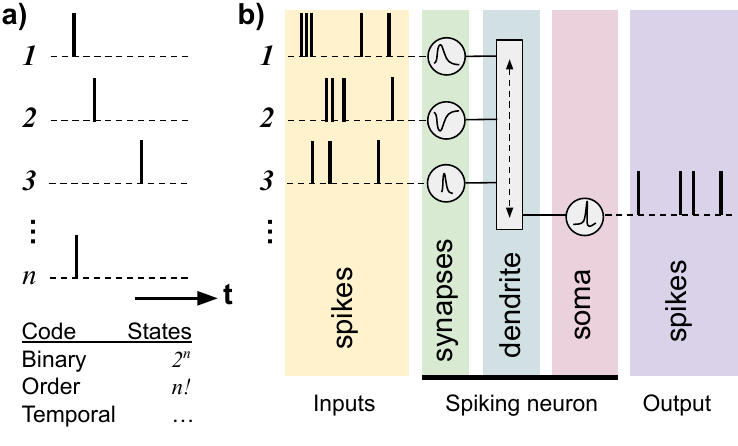}
    \caption{\emph{Spike-based encoding (a) and processing of information using a spiking neuron unit (b).
    Spikes are asynchronous binary events used to represent detector hits and neuron activations.
    By encoding information in a succession of spikes, such as their order or temporal distance, non-binary modes of information coding can be used.
    By leveraging physical phenomena for integration (in synapses) and processing (in dendrites and the soma of the neuron) of spike codes the energy efficiency and latency can be improved versus ordinary logic information processing~\cite{mehonic2024review}, in particular when the output precision is limited by noise~\cite{verhelst2015analogdigital}.}
}
    
    \label{fig:snn-scheme}
\end{figure}

The modeling of spiking neurons within SNNs typically involves integrating and firing action potentials represented as spikes, with each neuron having a membrane potential that rises in response to synaptic inputs, see Fig.~\ref{fig:snn-scheme}. When this potential surpasses a threshold, or the dynamical system reaches an unstable fixed point, the neuron emits one or several spikes that are transmitted to connected neurons (with some delays). Various mathematical models, such as the Leaky Integrate-and-Fire (LIF) model and the (adaptive) exponential integrate-and-fire model, are commonly used to describe this process. These models vary in complexity but share the core concept that the timing of incoming signals drives neuronal activity.

In neuromorphic computing, SNNs are used to model neural networks and brain-like algorithms in a biologically plausible way, taking advantage of the temporal coding of spikes, which makes them suitable for applications in spatiotemporal pattern recognition, sensor fusion, and real-time signal processing tasks that are often central to particle detection and analysis. By leveraging SNNs, neuromorphic systems can process high-dimensional events with complex asynchronous temporal structures.

In the context of particle detectors, SNNs offer significant potential for sparse encoding and processing of detector events. Their ability to process time-series data, such as signals from detectors that change over time or are influenced by various factors, could lead to more efficient, robust, and low-power systems for real-time dimension reduction, triggering, and event classification. As the field of neuromorphic computing continues to evolve, SNNs can have a pivotal role in revolutionizing how particle detection systems process information, making them an area of active research within both the neuromorphic and particle physics communities.

\subsection{Neuromorphic Computing in High-Energy Physics} 

Particle physics has a long history of synergy with computer science developments. Some of the applications required to study subnuclear matter and interactions, in fact, often require access to extensive computing resources, innovative algorithms, and specialized computing hardware. 

Given the above trend, it is no surprise that in recent years particle physics experiments have extensively adopted deep learning technologies, integrating them in their data analysis workflows~\cite{aiforhepbook}. In parallel, a number of efforts have been directed toward the integration of neural networks and other machine learning models in online data acquisition~\cite{Torino2021FPGAbasedDL}. Quantum Computing (QC) developments are also being followed closely by the HEP community, with a view to offering specific use cases where QC may provide successful encoding and implementation of the computing tasks~\cite{QC4HEP}.

A similar trajectory is now starting to become apparent for the Neuromorphic Computing (NC) paradigm. While the specific points of strength of NC technologies over traditional digital computing are not necessarily aligned with the most pressing demands of particle physics experiments in design or commissioning, there are situations where NC can provide advantageous alternative solutions that open a view into entirely different design concepts.
In this study, we focus on the Compact Muon Solenoid (CMS) Phase 2 detector at the High-Luminosity Large Hadron Collider (HL-LHC) as a case study to demonstrate the potential of neuromorphic computing in HEP. The detector's high pileup conditions, with an average of 200 simultaneous collisions per bunch crossing, present a significant computational challenge for real-time track reconstruction. By leveraging SNNs, we aim to address these challenges while maintaining high efficiency and low fake rates.

The structure of this article is as follows. In Sec.~2 we summarize how particle trajectories are identified and measured by the CMS tracking system, which is the use case on which we focus our attention in our study. In Sec.~3 we describe the SNN model we employ to demonstrate how a neuromorphic system may identify patterns of hits left by charged particles in silicon sensors, in a unsupervised way. We discuss the data samples we generated for our study in Sec.~4. In Sec.~5 we describe the tuning of hyperparameters of the SNN model, performed with a two-staged approach also employing a genetic algorithm. We detail our results in Sec.~6, and offer some concluding remarks in Sec.~7.

\section{Track Reconstruction with the CMS Phase2 experiment at HL-LHC} 

\begin{figure}[!h]
    \centering
    \begin{minipage}[b]{0.8\textwidth}
        \centering
        \includegraphics[width=\textwidth]{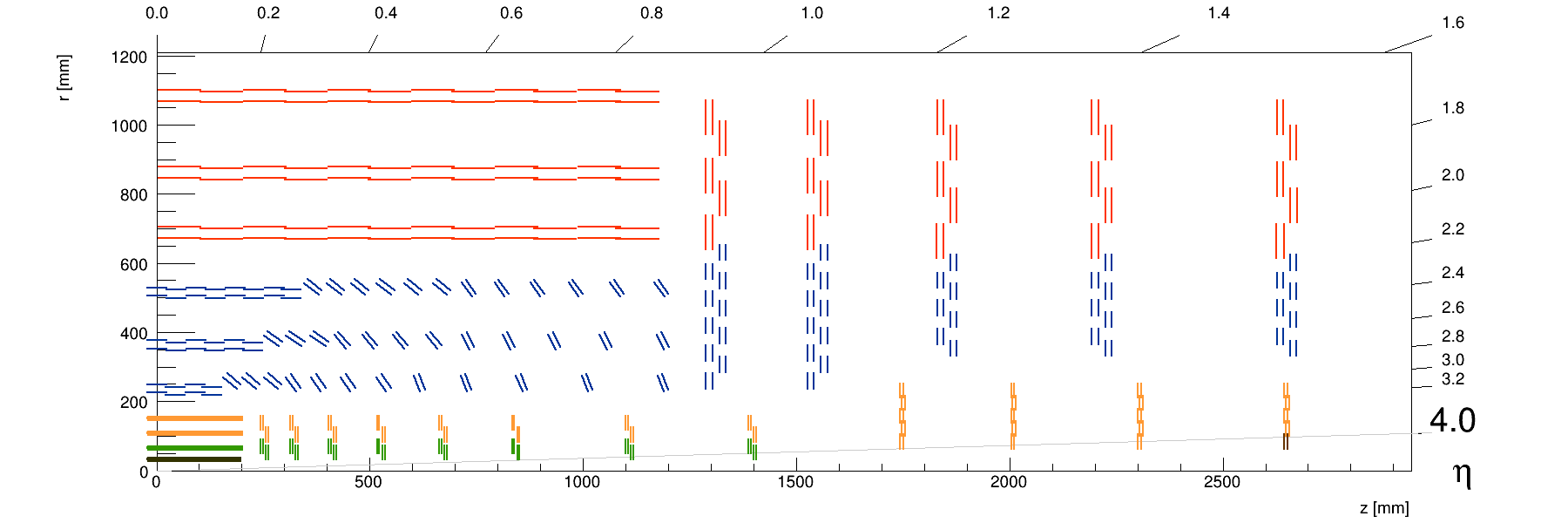}
        \caption{\emph{Layout of the silicon sensors in one sector of the Phase-2 CMS tracker~\cite{cms_tracker_layout}.}}
        \label{fig:layout_tracker_phase_2}
    \end{minipage} \hfill
\end{figure}

The Phase 2 CMS detector incorporates significant upgrades to its tracking system to meet the challenges of the High-Luminosity LHC (HL-LHC)~\cite{cms:TDR-014}. Figure \ref{fig:layout_tracker_phase_2} illustrates the layout of the silicon sensors within a single sector of the Phase-2 CMS tracker.

The redesigned all-silicon tracker provides exceptional spatial resolution and introduces timing information to improve particle trajectory reconstruction in high pileup environments, with an average of 200 simultaneous proton-proton collisions per bunch crossing.

The tracking system employs a sophisticated iterative reconstruction algorithm to identify charged particle trajectories, which ensures robust track reconstruction under the challenging conditions of HL-LHC:
\begin{itemize}
\item Seeding: Track seeds are formed using hits from the detector, focusing on high-efficiency seed generation even in high-density environments.
\item Trajectory Building: Using the Kalman filter, the algorithm extends the seed through the detector layers, accounting for multiple scattering and energy loss.
\item Fitting: A final fit refines the trajectory, providing precise momentum, charge, and vertex information.
\end{itemize}

This iterative approach prioritizes high-momentum and prompt tracks early and handles low-momentum tracks and complex patterns in subsequent iterations, improving efficiency and fake rejection.
The following key performance metrics emphasize the capabilities of the track reconstruction process and its impact on the overall experimental performance.
\begin{itemize}

\item Track Efficiency: The track reconstruction algorithm achieves exceptionally high efficiency. For charged particles with transverse momentum ($p_T > 1\ \text{GeV}$) it exceeds 99\% in the central region ($|\eta| < 1.5$) and it still remains above 95\% in the forward region ($1.5 < |\eta| < 4.0$).
\item Transverse Momentum Resolution: The Phase-2 track reconstruction achieves a transverse momentum resolution of better than 1-2\% for high-$p_T$ tracks in the central region ($|\eta| < 1.5$), which is critical for high-precision measurements of particle momentum. Although the resolution slightly degrades in the forward region due to the increased material and multiple scattering effects, it remains within acceptable bounds for reliable track reconstruction across the entire detector.
\item Fake Track Rate: The fraction of reconstructed tracks that do not correspond to real particle trajectories, often arising from noise, overlapping hits, or mis-reconstruction, is expected to be 0.5\% to 1\% for tracks with $p_T > 1\ \text{GeV}$.
\end{itemize}

Efficient computing timing for track reconstruction in the CMS Phase-2 detector is crucial for handling the high-luminosity, high-pileup conditions expected at the HL-LHC. Key performance optimizations include multi-threading, GPU acceleration, and potentially machine learning algorithms to handle the increased event complexity while maintaining high precision and efficiency.

\section{Spiking Neural Network Model for Particle Tracking} 
\subsection{SNN Architecture}
The model employed in this work builds upon the SNN architecture proposed by Masquelier {\it et al.}~\cite{Masquelier2009}, specifically designed to perform spike-timing-based learning via Spike-Timing-Dependent Plasticity (STDP). This architecture integrates Leaky Integrate-and-Fire (LIF) neurons to achieve the recognition of complex spatiotemporal patterns in a noisy environment, facilitated through an unsupervised learning process. 

\begin{figure}[!h]
    \centering
    \includegraphics[width=0.8\linewidth]{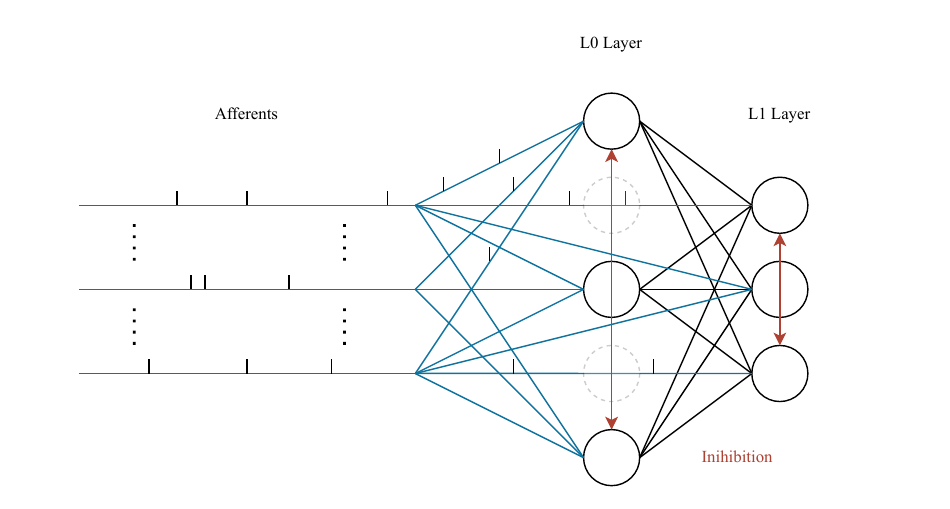}
    \caption{\emph{Sketch of the Network Architecture. In the final network presented in Sec.~\ref{sec:final_results} there are 10 afferents, $N_{L_0}=6$ neurons in layer $L_0$ and $N_{L_1}=6$ neurons in layer $L_1$.}}
    \label{fig:network_architecture}
\end{figure}

Unlike the model proposed by Masquelier et al., the architecture employed in this work consists of two primary layers, $L_0$ and $L_1$, which are densely connected to afferents serving as channels for input signals. Each afferent corresponds to a specific input source, which introduces spikes into the network, simulating a diverse array of sensory inputs. Each of the $N_{L_0}$ $\left (N_{L_1} \right )$ neurons of the layer $L_0$ $\left (L_1 \right )$ are characterized by the activation threshold $T_0$ $\left (T_1 \right )$. Furthermore, each synapse $j$ linked to an afferent has a synaptic delay $d_j$, which are additional degrees of freedom that our model exploits with a novel learning algorithm presented in Sec.~\ref{sec:DSTDP}. Similar unsupervised learning rules for the synaptic delays have been studied by other works like  \cite{hammouamri2023learningdelaysspikingneural}~\cite{nadafian2020bioplausibleunsuperviseddelaylearning} ~\cite{hazan2022memorytemporaldelaysweightless}. A simplified scheme of the network architecture employed for this study is shown in Figure \ref{fig:network_architecture}.

\subsection{Initialization of the Synaptic Weights and Delays}

The initial synaptic weights are drawn from a Gaussian distribution with mean $\mu=1$ and standard deviation $\sigma = 2 /\sqrt{N_\text{afferents}} = 2 / \sqrt{10}$. Then, they are normalized so that their sum is unitary.
Synaptic delays are initialized to random values within the range $[d_\text{max}/ 2-\Delta, d_\text{max}/2 +\Delta]$ where $\Delta$ is a hyperparameter set to ensure sufficient temporal spread and $d_\text{max}$ is the maximum value allowed for the synaptic delay. These initial values were chosen to provide a diverse starting point for the unsupervised learning of spatio-temporal patterns.

\begin{figure}[!h]
    \centering
    \begin{minipage}{0.9\linewidth}
        \centering
        \includegraphics[width=\linewidth]{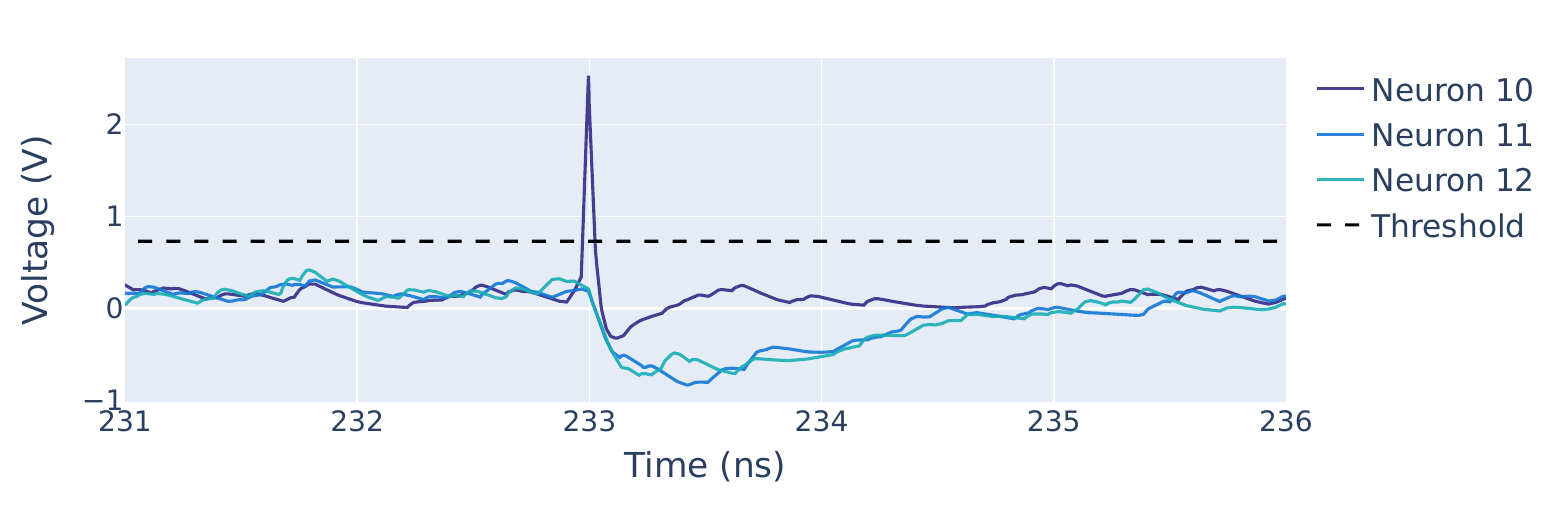}
        \label{fig:firing-event}
    \end{minipage}
    \vspace{-0.75cm}
    \begin{minipage}{0.9\linewidth}
        \centering
        \vspace{-0.75cm}\includegraphics[width=\linewidth]{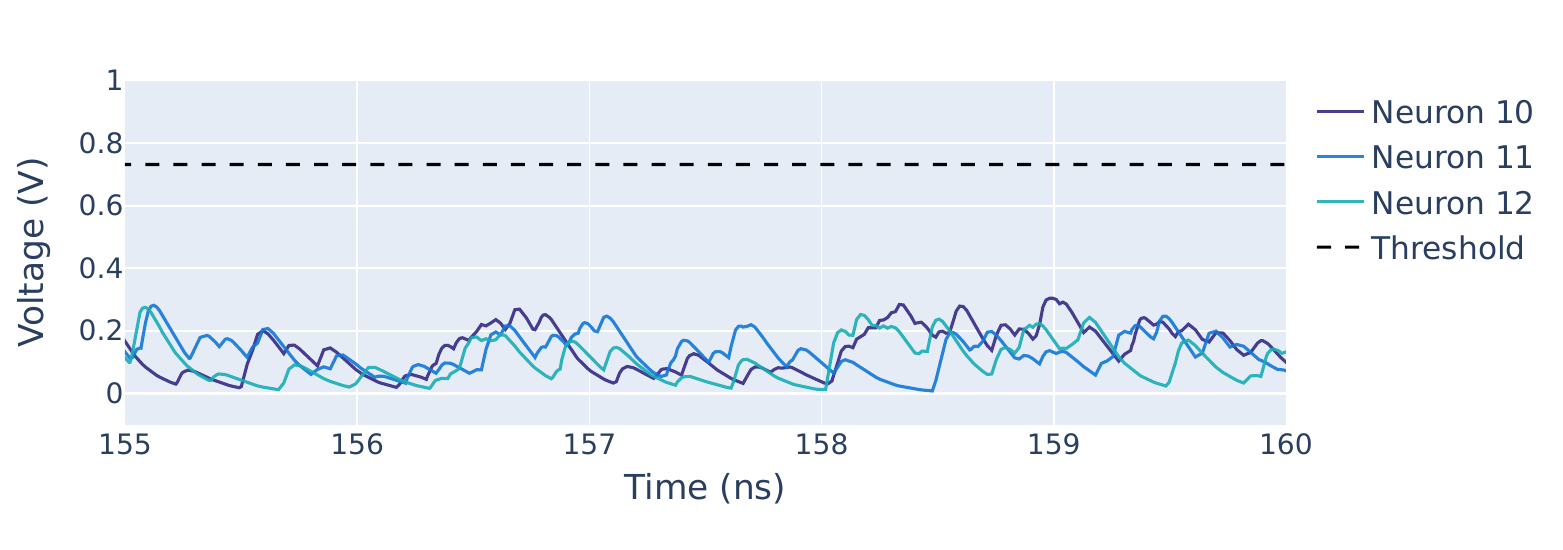}
        \label{fig:non-firing-event}
    \end{minipage}
    
    \caption{\emph{Comparison of membrane potential evolution in firing and non-firing events. 
    In the top plot, neuron 10 surpasses the firing threshold at $t \approx 143.7 \, \text{ns}$, experiencing a sharp increase in potential followed by a reset governed by the reset potential $\eta(t - t_i)$. An entire event is $25 \, \text{ns}$.
    The firing neuron also induces an IPSP $\mu(t - t_k)$ in the other neurons, suppressing their activation and reinforcing the competition mechanism. 
    In the bottom plot, neurons exhibit sub-threshold oscillations but do not fire, indicating that the input signals were insufficient to reach the activation threshold.}}
    \label{fig:firing-vs-non-firing}
\end{figure}

\subsection{Numerical simulation of the Neuron Potentials}
The postsynaptic potential (EPSP) produced by an incoming spike at time $t_j$ is computed as:
\begin{equation}
    \epsilon(t - t_j) = K \cdot \left[ \exp \left( -\frac{t - t_j}{\tau_m} \right) - \exp \left( -\frac{t - t_j}{\tau_s} \right) \right] \cdot \theta(t - t_j)
\end{equation}
Here, $\tau_m$ and $\tau_s$ represent the membrane and synaptic time constants, respectively, and $\theta(\cdot)$ is a unit step function. $K$ is a scaling constant such that the peak of the potential is set to 1: $\max_t{\epsilon(t)} = \epsilon_{t_{\text{max}}}=1$. This potential change captures the dynamics of how an individual synapse contributes to the membrane potential of a neuron upon receiving an input spike. After reaching a certain threshold T, the neuron fires. 
Upon firing at time $t_i$, the neuron experiences a reset in its membrane potential, described by:
\begin{equation}
    \eta(t - t_i) = T \cdot \left[ K_1 \cdot \exp \left( -\frac{t - t_i}{\tau_m} \right) - K_2 \cdot \left( \exp \left( -\frac{t - t_i}{\tau_m} \right) - \exp \left( -\frac{t - t_i}{\tau_s} \right) \right) \right] \cdot \theta(t - t_i)
\end{equation}
with $K_1$ and $K_2$ acting as constants that dictate the post-firing behavior.
Inhibitory Postsynaptic Potentials (IPSPs) are incorporated to introduce competition between neurons within the network. When a neuron fires at time $t_k$, the resulting inhibitory potential it sends to its neighbors is expressed as:
\begin{equation}
    \mu(t - t_k) = -\alpha \cdot T \cdot \epsilon(K_\mu(t - t_k))
\end{equation}
where $\alpha$ is a coefficient representing the strength of inhibition and $K_\mu$ adjusts its temporal extension. This competitive interaction among neurons promotes selective firing, preventing over-activation of the network.
As a consequence of the IPSPs, a 'Winner-Takes-All' competition mechanism emerges, where neurons compete to respond to input patterns. This process drives specialization, with neurons becoming increasingly tuned to distinct features of the input streams.
Thus, at a given point in time, the membrane potential of a neuron is given by:
\begin{equation}
    p(t) = \eta(t - t_i) + \sum_{j | (t_j+d_j) > t_i} w_j \cdot \epsilon(t - (t_j+d_j)) + \sum_{k | t_k > t_i} \mu(t - t_k)
\end{equation}
with $w_j$ and $d_j$ representing the synaptic weight and delay. This defines the model of an LIF neuron.
The main advantage of the model is that the calculations on potentials are performed only when a neuron receives or emits spikes, making it extremely efficient compared to solving a differential equation that describes potential dynamics.

\subsection{Modified STDP for Unsupervised Synaptic Delay Learning}
\label{sec:DSTDP}
STDP is a learning mechanism inspired by biological processes that modifies synaptic weights depending on the timing difference between incoming spikes and neuronal activation~\cite{Song2000}~\cite{Bi2001}. Within SNNs, STDP provides a framework for unsupervised learning. The Hebbian STDP model in~\cite{Masquelier2009} is formulated as follows: 
\begin{definition}[Hebbian STDP rule for Synaptic Weights]
   \begin{equation*}
    \Delta w_j = 
    \begin{cases} 
        a^+ \cdot \exp \left( \frac{t_j - t_i}{\tau^+} \right) & \text{if } t_j \leq t_i \quad \Rightarrow \text{ Synaptic Long-Term Potentiation
} \\
        -a^- \cdot \exp \left( -\frac{t_j - t_i}{\tau^-} \right) & \text{if } t_j > t_i \quad \Rightarrow \text{ Synaptic Long-Term 
 Depression}
    \end{cases}
\end{equation*} 
Where $t_j$ denotes the pre-synaptic spike arrival time, $t_i$ is the neuron activation time, $a^+$ and $a^-$ define the maximum weight update, and $\tau^+$ and $\tau^-$ are constants that control the duration of the time window in which spikes lead to either synaptic potentiation or depression. 
\end{definition}
According to this rule, if an input spike arrives prior to neuronal activation, the synaptic weight increases, reinforcing the causality relation; conversely, if it arrives afterward, the weight decreases.

\begin{figure}[!h]
    \centering
    \includegraphics[width=0.6\linewidth]{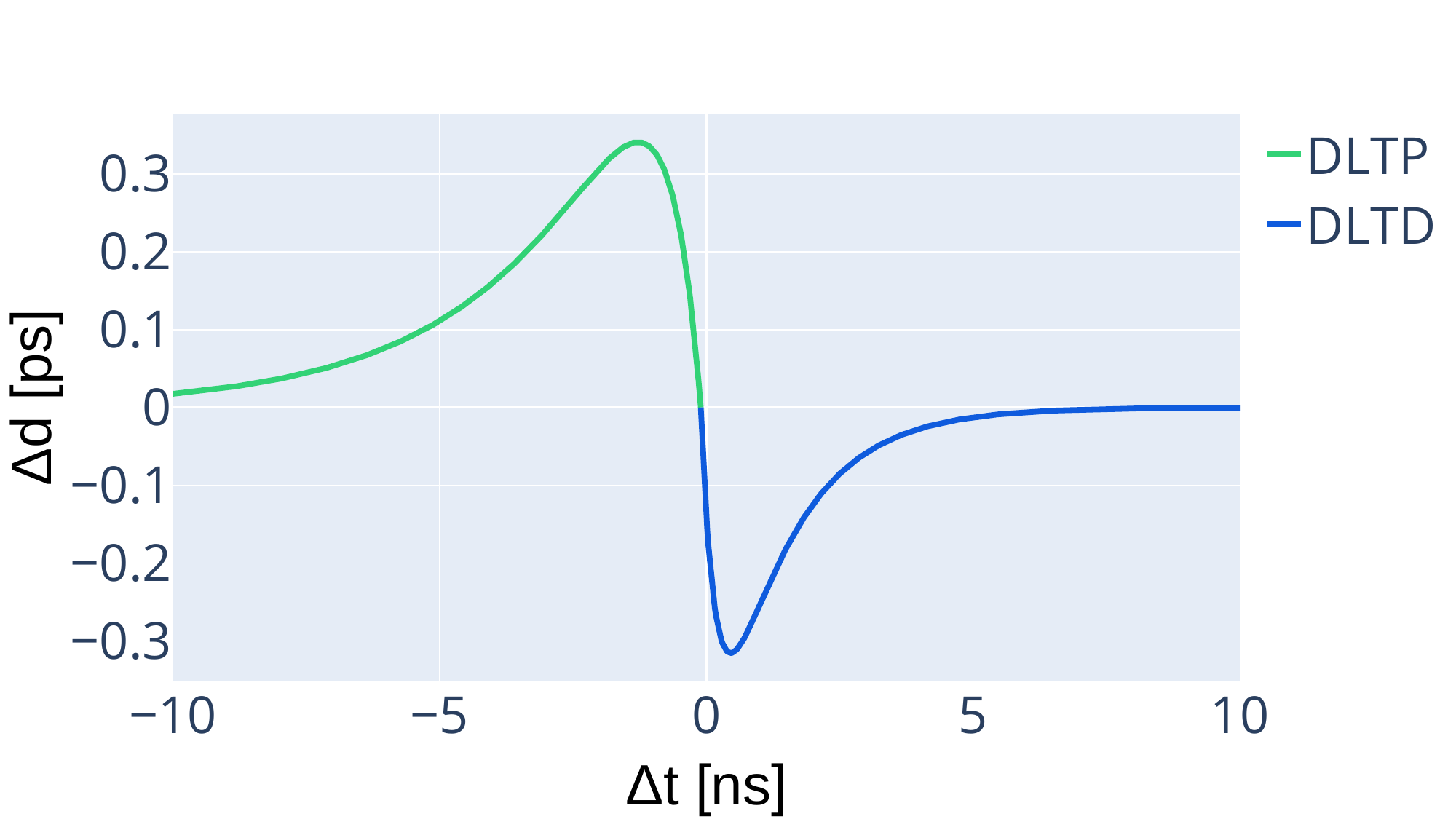}
    \caption{\emph{Model of spike-time-dependent plasticity of connection delays implementing a modified Hebbian learning concept. The x-axis denotes the time difference, $\Delta t =t_j - t_i+t_{\text{max}}$, and the y-axis, $\Delta d$, is the corresponding update of the delay. The parameters defining this unsupervised learning rule were found by the genetic algorithm.}}
    \label{fig:stdp-delay}
\end{figure}

This rule alone has proven insufficient for the purposes of this work, as it fails to lead neurons to specialize in recognizing distinct patterns (see Sec.~\ref{sec:weights-stdp-results}).
To address this issue, we propose a modified version of this rule (see Fig.~\ref{fig:stdp-delay}) specifically for learning synaptic delays instead, by incorporating: 
\begin{itemize}
    \item \textbf{Delay Potentiation (DLTP):} If a pre-synaptic spike occurs before the neuron's activation, the delay is increased, effectively delaying the pre-synaptic spike.
    \item \textbf{Delay Depression (DLTD):} If a pre-synaptic spike occurs after the neuron's activation, the delay is decreased, advancing the pre-synaptic spike.
\end{itemize}
Through this mechanism, neurons can adjust connection delays (heretofore addressed as ``synaptic delays'') to minimize the temporal pattern length, increasing the activation likelihood. Analytically, this rule is defined as follows:

\begin{definition}[STDP rule for synaptic delays]
\begin{equation*}
    \Delta d_j =
    \begin{cases} 
        d_+ \left [ \exp \left( \frac{t_j - t_i+t_{\text{max}}} {\tau_{d_+}} \right ) -  \exp \left( \frac{t_j - t_i + t_{\text{max}}}{\tau_{d_+}^{'}} \right )\right] & \text{if } t_j \leq t_i - t_{\text{max}}\, \Rightarrow\text{DLTP},  \\-d_- \left [ \exp \left( \frac{t_i - t_j - t_{\text{max}}}{\tau_{d_-}} \right) - \exp \left( \frac{t_i - t_j - t_{\text{max}}}{\tau_{d_-}^{'}} \right) \right] & \text{if } t_j > t_i - t_{\text{max}}\, \Rightarrow\text{DLTD}, \\
        0  & \text{if synapse j is linking two neurons}
        \end{cases}
    \label{STDP_delay}
\end{equation*}
Where $t_j$ denotes the pre-synaptic spike arrival time, $t_i$ represents the neuron's activation time, $t_{\text{max}}$ indicates the EPSP signal's peak time, $d_+$ and $d_-$ are the learning rates, and $\tau_{d_+}$, $\tau_{d_+}^{'}$, $\tau_{d_-}$, and $\tau_{d_-}^{'}$ are the time constants for potentiation and depression, respectively. 
The rule updates the delay of synapses linking afferents to neurons.
\end{definition}
As a regularization method, the sum of the synaptic delays of each neuron remains unchanged throughout the process, with a renormalization step conducted after each update. Furthermore, delays are clipped in a definite range $d_j\in[0,d_\text{max}]$.

\FloatBarrier

\subsection{Spatiotemporal Information Encoding of Detector Hits}

\begin{wrapfigure}{l}{0.4\textwidth} 
    \centering
    \includegraphics[width=0.38\textwidth]{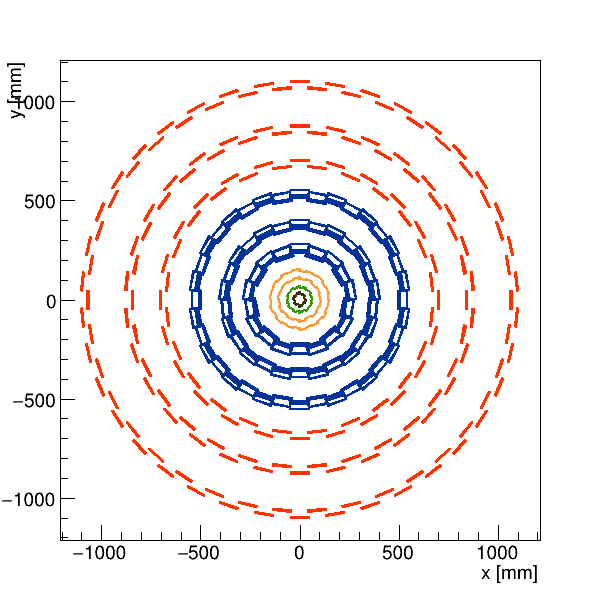} 
    \caption{\emph{Layout of the silicon sensors in the Barrel of the Phase-2 CMS tracker in the transverse plane~\cite{cms_tracker_layout}.}}
    \label{fig:barrel_layout}
\end{wrapfigure}

The challenge of encoding information involves determining how to map the three-dimensional coordinates $(r,\phi,\eta)$ of hits recorded by the tracker into input spikes $(a,t)$ received at a specific time $t$ from a specific afferent $a$. Given our objective to distinguish particles from noise and quantify their transverse momenta, we initially concentrate on the radial coordinate and azimuthal angle, postponing the integration of the third dimension for future investigations. This approach effectively considers events as projected onto the transverse plane. Furthermore, our current focus is limited to events involving only the Inner Tracker (IT) and Outer Tracker (OT) barrel, deferring any inclusion of the endcaps for later expansions. The geometry corresponds to what is displayed in Fig.~\ref{fig:barrel_layout}. Specifically, we assign to each tracking layer a unique afferent, and scanning the tracker counterclockwise beginning from the positive half of the x-axis. Define the angular reading speed as $\omega = 2\pi\cdot f$, with $f=40\ \text{MHz}$ representing the LHC event rate. The $j^{th}$ hit $(r_j,\phi_j)$ is thus encoded into a spike $(a_j,t_j)$ where $a_j$ corresponds to the afferent related to its tracking layer and $t_j = \frac{\phi_j+2\pi\cdot\theta(-\phi_j)}{\omega}$ represents the arrival time. Consequently, we establish a mapping $r_j \Rightarrow a_j$ and $\phi_j \Rightarrow t_j$ that remains monotonic and continuous for signal tracks, except at the discontinuity transitioning between $\phi = 2\pi\ \text{rad}$ and $\phi=0$.

The implemented signal encoding method faces challenges with edge cases near the $\phi = 2\pi \ \text{rad}$ to $\phi = 0$ transition, where trajectories overstep the boundary of the encoding region. To resolve this, the encoding framework is expanded to include an auxiliary region $\phi \in [0, \delta]$ with $\delta = 0.7 \ \text{rad}$. This covers the maximum anticipated angular discrepancy between the closest and farthest hit for the lowest momentum track ($1 \, \text{GeV}$), ensuring continuous signal representation and preserving monotonicity, albeit introducing minimal redundancy.

\begin{figure}[!h]
    \centering
\includegraphics[width=\linewidth]{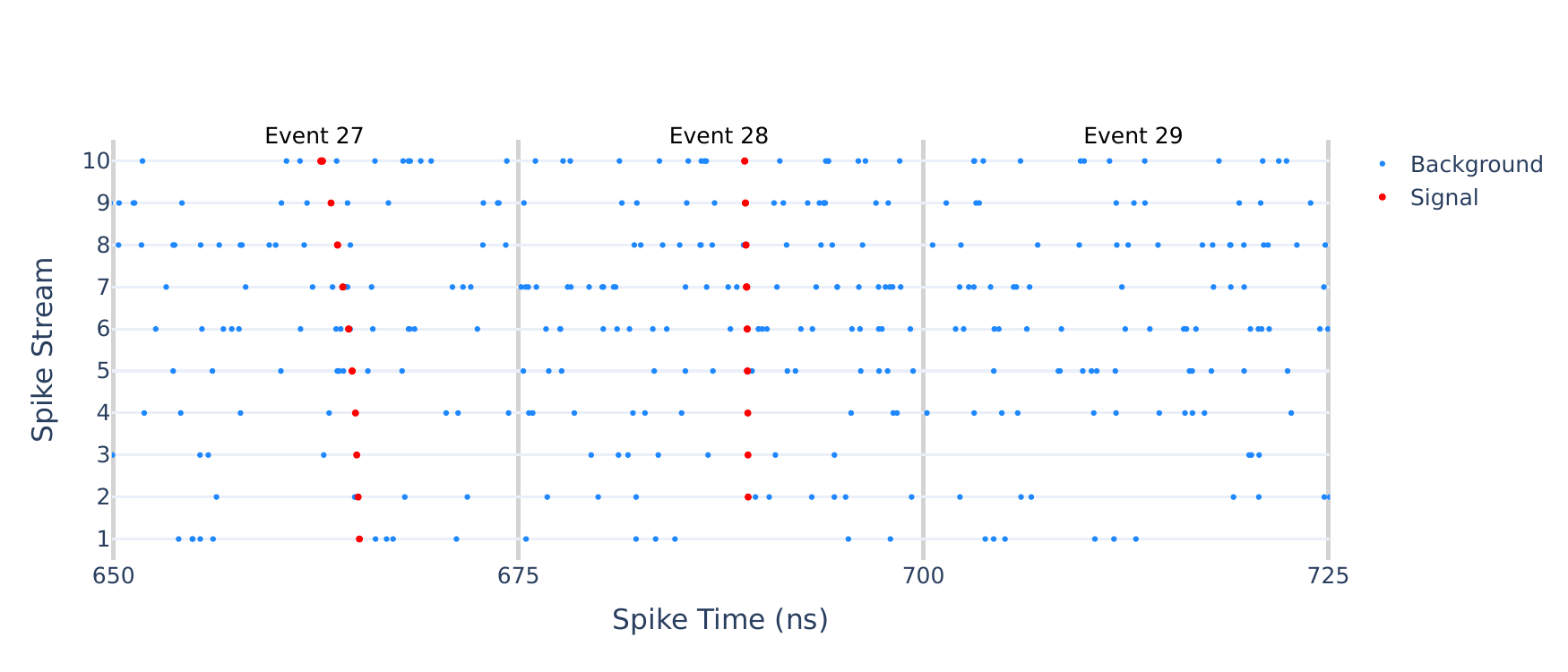}
    \caption{\emph{Spatiotemporal spike patterns. The x-axis represents encoding time, while the y-axis designates each afferent. Each dot symbolizes an incoming spike from the associated afferent. Blue dots denote spikes associated with clusters of background, whereas red dots represent spikes associated with clusters of signal}}
    \label{fig:signal-coding}
\end{figure}
To validate the proposed SNN model, a carefully simulated dataset is essential. This dataset must capture the spatiotemporal patterns of charged particle trajectories while incorporating realistic noise and detector geometry. As illustrated in Figure \ref{fig:signal-coding}, the dataset exhibits distinct clusters of signal spikes amid background noise, effectively preserving the event-based nature of the encoding. In the following section, we describe the dataset preparation process and its role in training and evaluating the SNN. In the following section, we describe the dataset preparation process and its role in training and evaluating the SNN.
\FloatBarrier

\section {Simulated Dataset for Training and Validation} 
The dataset employed for training the SNN was derived from Monte Carlo simulations based on the Phase 2 detector geometry. Half of the events contain just random noise, while the other half were generated assuming the production of a single particle (either a muon or an antimuon) without pile-up conditions. The kinematic properties of the particles were uniformly distributed in azimuthal angle $\phi \in [-\pi, \pi]$ and pseudorapidity $\eta \in [-1, 1]$, focusing on the Barrel region while excluding inclined disks and detectors. Three transverse momentum classes were considered, corresponding to $p_T \in \{1, 3, 10\} \, \mathrm{GeV}$.

The spatial density of hits $\rho(\vec{x})$ is modeled as $\rho(\vec{x}) \propto F(\vec{x}) \cdot G(\vec{x})$, where $F(\vec{x})$ represents the structural features of the tracker (vanishing in regions without active sensors) and $G(\vec{x})$ accounts for the trajectory geometry. Given that the primary events were uniformly distributed in $\phi$ and $\eta$, the factor $G(\vec{x})$ is proportional to $1/r$. The generation of the background hits for network training and testing has been done following a custom generation process, extracting random signal hits through an inversion-based sampling algorithm:
\begin{enumerate}
    \item Hits from the Monte Carlo simulations were collected into a reference set.
    \item Each hit was assigned a weight equal to 1.
    \item A cumulative distribution function $P(i) = \sum_{j \leq i} r_j$ was computed for the reference set.
    \item A random value $x$ was drawn uniformly in $[0, P_\text{max}]$, where $P_\text{max}$ is the maximum value of $P(i)$.
    \item The corresponding hit index $i_\text{hit}$ was determined by inverting $P(i)$.
    \item Steps 4 and 5 were repeated until the desired number of noise hits was obtained.
\end{enumerate}

\begin{figure}[!h]
    \centering
    \begin{subfigure}[b]{0.45\linewidth}
        \centering
        \includegraphics[width=\linewidth]{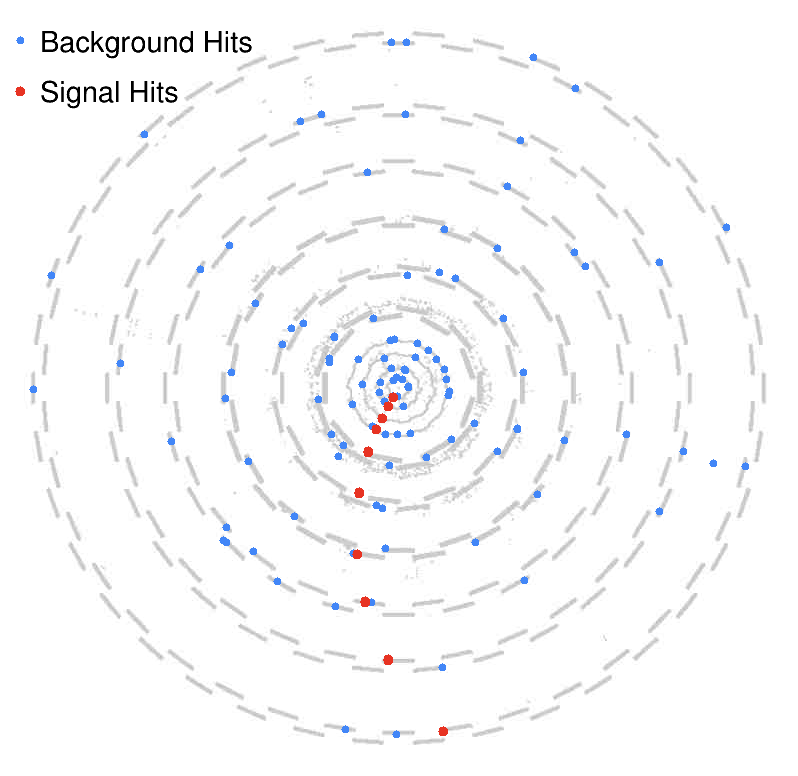}
        \caption{}
        \label{fig:signal-100}
    \end{subfigure}
    \begin{subfigure}[b]{0.45\linewidth}
        \centering
        \includegraphics[width=\linewidth]{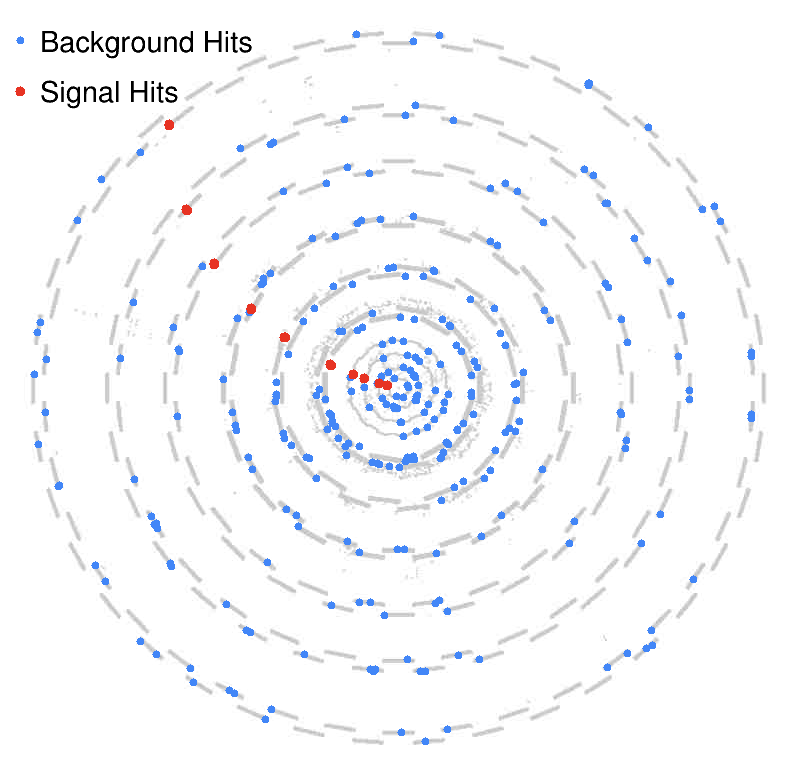}
        \caption{}
        \label{fig:signal-300}
    \end{subfigure}
    \caption{\emph{Comparison of transverse plane projections of events with different noise levels. The left panel shows an event containing 100 noise clusters, and the right panel shows an event with 300 noise clusters. Red points represent clusters associated with the main event, while blue points indicate background noise.}}
    \label{fig:signal_comparison}
\end{figure}

\begin{figure}[!h]
    \centering
    \begin{minipage}[b]{0.45\textwidth}
        \centering
        \includegraphics[width=\textwidth]{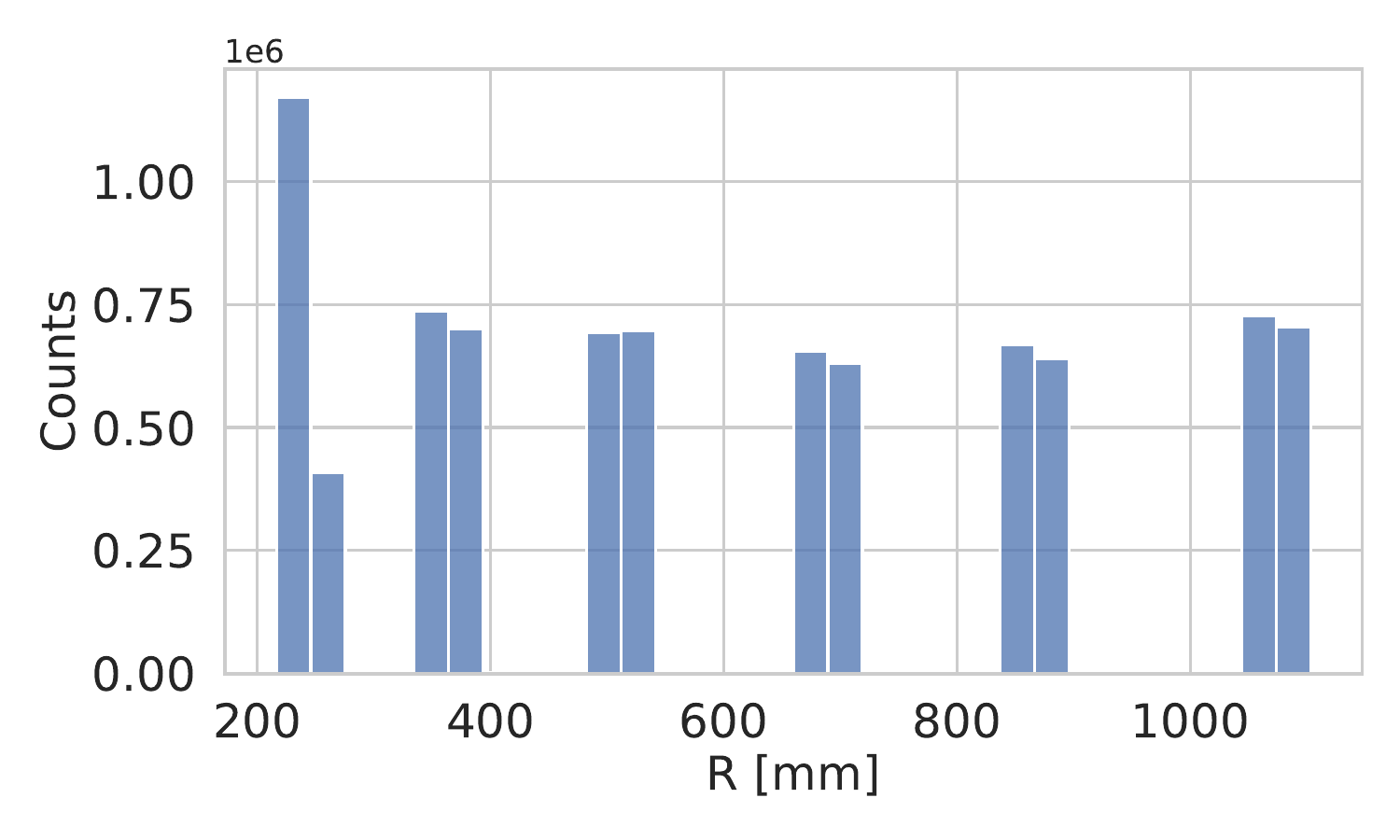}
        \caption{\emph{Distribution of the hits in $R$, the radial distance in the transverse plane from the detector's origin.}}
        \label{fig:dist_hist_R}
    \end{minipage}
    \hfill
    \begin{minipage}[b]{0.45\textwidth}
        \centering
        \includegraphics[width=\textwidth]{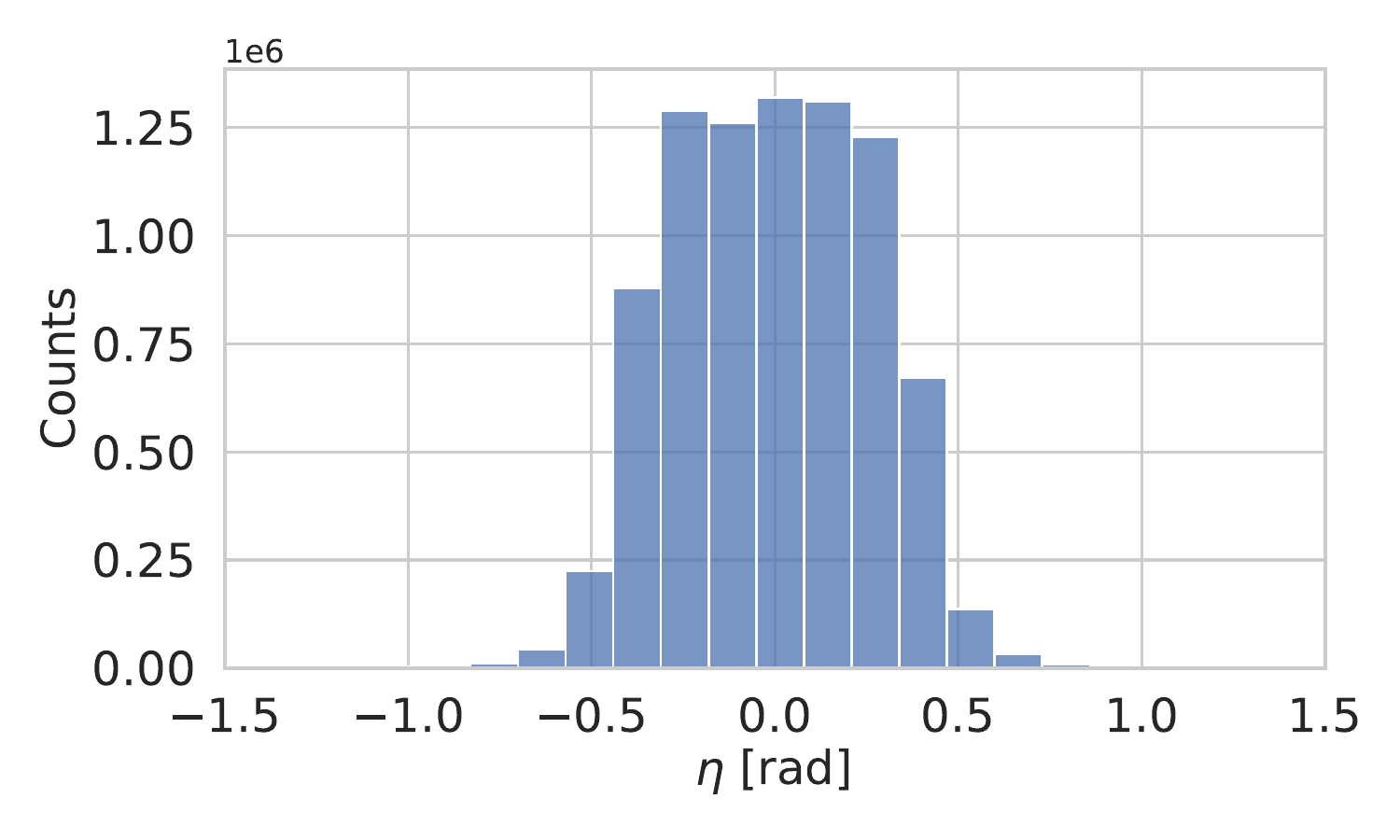}
        \caption{\emph{Distribution of $\eta$, the pseudorapidity, indicating the spread of clusters along the beam axis direction.}}
        \label{fig:dist_hist_eta}
    \end{minipage}
    
    \vspace{0.5cm}
    \begin{minipage}[b]{0.45\textwidth}
        \centering
        \includegraphics[width=\textwidth]{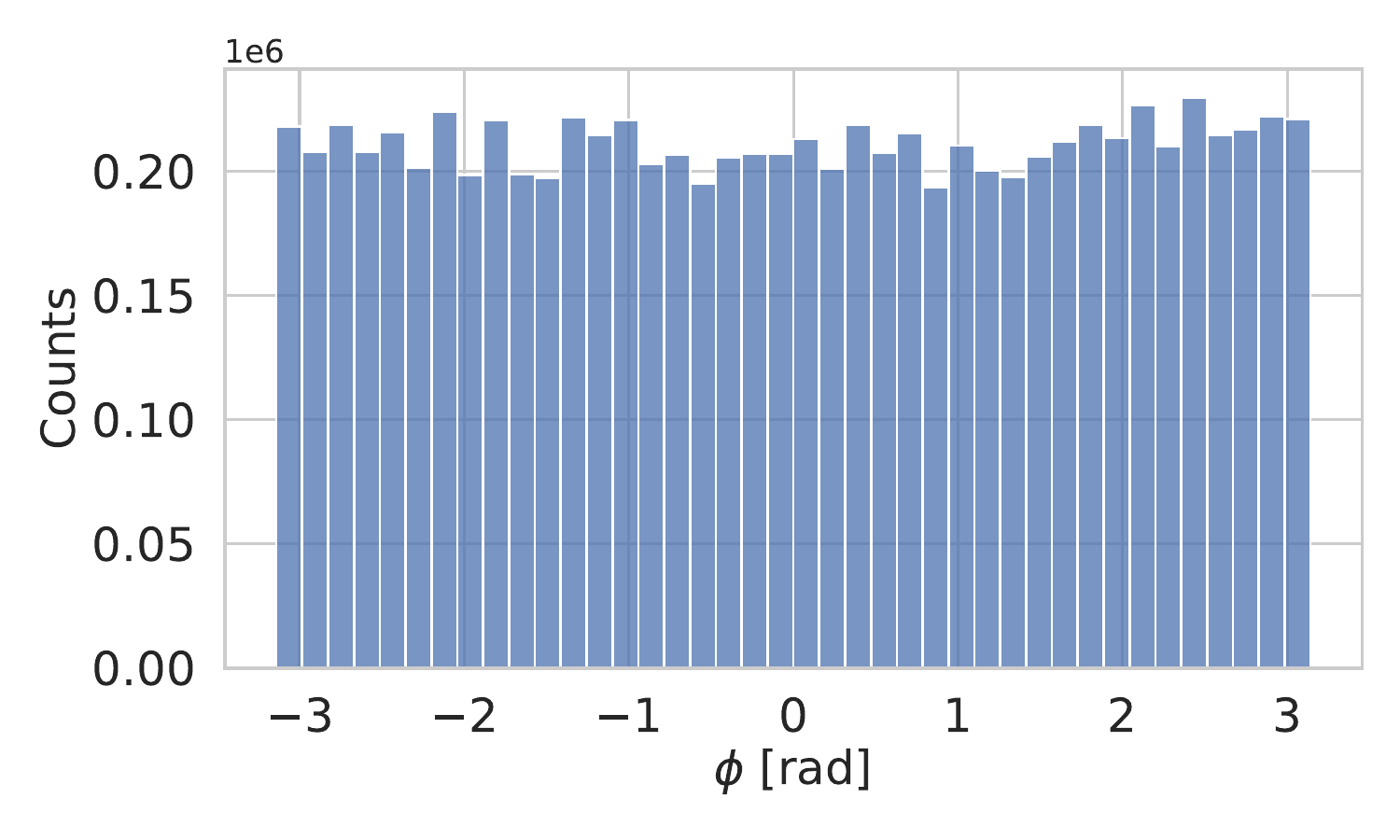}
        \caption{\emph{Distribution of $\phi$, the azimuthal angle, showing the uniformity of clusters around the detector transverse plane.}}
        \label{fig:dist_hist_phi}
    \end{minipage}
\end{figure}

This approach maintains the spatial consistency of the noise distribution with respect to the detector geometry. The number of hits of noise in an event is supposed to follow a Poisson distribution with a given average $\langle{N_{\text{bg}}\rangle}$. Figure \ref{fig:signal-100} illustrates a representative event where $\langle N_{\text{bg}} \rangle = 100$ noise clusters, generated using this method, are overlaid onto the primary signal. In contrast, Figure \ref{fig:signal-300} shows the event with 300 background noise clusters. This dataset preparation strategy ensures a balanced and realistic input for network training, preserving both signal fidelity and background complexity. Figures \ref{fig:dist_hist_R}, \ref{fig:dist_hist_eta},\ref{fig:dist_hist_phi} show the distributions of the generated datasets. 
The next step involves training and optimizing the SNN to handle high-noise environments effectively. This requires fine-tuning hyperparameters such as synaptic delays and learning rates to maximize efficiency and selectivity while minimizing fake rates. The following section describes our approach to hyperparameter optimization using genetic algorithms.

\FloatBarrier

\section{Methodology - Hyperparameter search and optimization} 

In this section we present the definitions of the utilities used to measure the performance of the SNN (\ref{sec:utilities}), discuss the problem of the specialization of the neurons (\ref{sec:utilities}) and how we tackled hyperparameter tuning via a genetic algorithm (\ref{sec:genetic_algorithm} and ~\ref{sec:Optimization-Workflow}).

\subsection{Utilities Definition}
\label{sec:utilities}

The index \( n \) represents the \( N_{\text{neurons}} \) neurons, while \( c \) denotes the index corresponding to the \( N_{\text{classes}} \) particle classes, using a notation where \( c = 0 \) refers to events containing only background hits. Alternatively, to specify a particular class, the notation \( c = (q, p_T) \) is used, where \( q \in \{-1, +1\} \) represents the charge state, and \( p_T \in \{1, 3, 10\}\ \text{GeV} \) corresponds to the transverse momentum. The symbol \( \epsilon \) serves as the index over the \( N_\epsilon \) events, while \( \epsilon_c \) specifically denotes events that include a track from a given class \( c \). To evaluate the functionality of the network and monitor its learning progression, the following definitions are introduced.

\begin{definition}[Neuron Activation Indicator Function]
    This function is used to identify events in which the neuron $n$ has been activated at least once.
    A neuron is considered ``activated at least once'' during an event if its membrane potential exceeds the firing threshold $T_0$ or $T_1$ within the duration of the event.
    \begin{equation*}
        1_n(\epsilon) = 
    \begin{cases}
        1 & \text{if the neuron $n$ has activated at least once during the event $\epsilon$} \\
        0 & \text{otherwise}
    \end{cases}
    \end{equation*}

\end{definition}

\begin{definition}[Network Activation Indicator Function]
    \begin{equation*}
        1(\epsilon) = 
    \begin{cases}
        1 & \text{if at least one neuron in the network has activated during event $\epsilon$} \\
        0 & \text{otherwise}
    \end{cases}
    \end{equation*}

\end{definition}

\begin{definition}[Acceptance per neuron per class] 
    Measures the fraction of events of class $c$ in which the neuron $n$ has been activated at least once.
    \begin{equation*}
        A_{n, c}=\frac{\sum_{\epsilon_c} 1_n( \epsilon_c)}{\sum_{\epsilon_c} 1}
    \end{equation*}
\end{definition}

\begin{definition}[Fake rate per neuron]
    Measures the fraction of events containing just hits of background in which the neuron $n$ has been activated at least once.
    \begin{equation*}
        F_{n} = A_{n, 0} = \frac{\sum_{\epsilon_0} 1_n( \epsilon_0)}{\sum_{\epsilon_0} 1}
    \end{equation*}
\end{definition}
\label{def:FakeRate_neuron}

\begin{definition}[Aggregate acceptance per class] 
    Measures the fraction of events of class $c$ in which at least one neuron in the network has been activated.
    \begin{equation*}
    A_{c}=\frac{\sum_{\epsilon_c} 1( \epsilon_c)}{\sum_{\epsilon_c} 1}
    \end{equation*}
\end{definition}

\begin{definition}[Aggregate fake rate]
    Measures the fraction of events containing just hits of background in which at least one neuron in the network has activated.
    \begin{equation*}
        F = A_{0} = \frac{\sum_{\epsilon_0} 1( \epsilon_0)}{\sum_{\epsilon_0} 1}
    \end{equation*}
\end{definition}
\label{def:FakeRate}
\begin{definition}[Selectivity of the network] 
     Selectivity quantifies the ability of the network to discriminate patterns and is derived from mutual information by comparing the distribution of activations across neurons and particle classes:
    \begin{equation*}
        S = \sum_{n,c} {P_{n,c}\cdot\log_2\left(\frac{P_{n,c} + \delta}{P_c\cdot P_n}\right)}
    \end{equation*}
    with $P_{n,c} = \frac{\sum_{\epsilon_c} 1_n( \epsilon_c)}{N_\epsilon}$,  $P_{n} = \sum_{c=1}^{N_{\text{classes}}}P_{n,c}$, $P_{c} = \sum_{n=1}^{N_{\text{neurons}}}P_{n,c}$ and $\delta \ll 1$ to avoid numerical instabilities.
\end{definition}
\label{def:selectivity}

\subsection{The Problem of Specializing Neurons}
\label{sec:weights-stdp-results}

\begin{figure}[!h]
    \centering
    \includegraphics[width=0.8\linewidth]{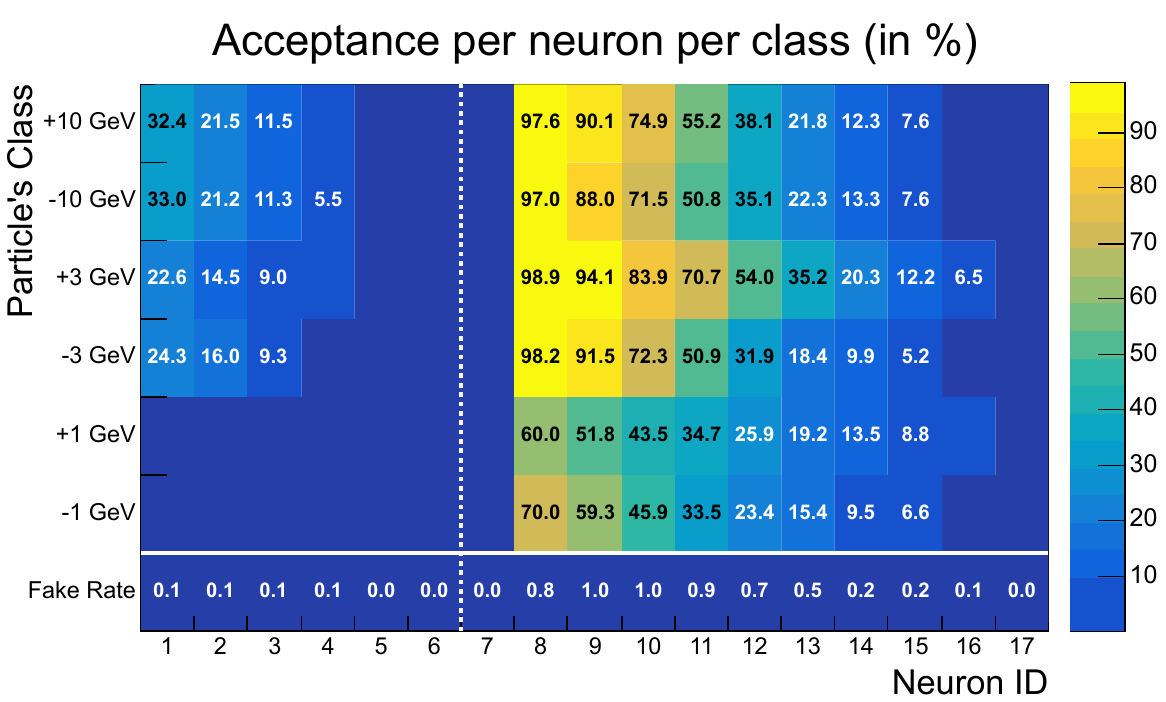}
    \caption{\emph{Heatmap of neuron activations across particle classes, with neurons represented on the x-axis and particle classes (characterized by charge and transverse momentum) on the y-axis. The color intensity indicates the acceptance rate, or the fraction of events in which the neuron was activated for a given class. The fake rate (false positives for noise-only events) is shown in the bottom row. This visualization demonstrates the lack of specialization of the neurons for specific classes, despite maintaining low false-positive rates.}}
    \label{fig:baseline-efficiencies}
\end{figure}

Initially, given the similarity of the task, the network was operated using the optimal parameters found in~\cite{Masquelier2009}, with all synaptic delays set to zero and weight learning enabled only. This configuration resulted in high acceptance rates (greater than 90\%) and low fake rates (below 5\%). For aggregated results, see Table~\ref{tab:aggregate_results_thesis}. However, with this configuration, the neurons did not achieve meaningful specialization, as shown in Fig.~\ref{fig:baseline-efficiencies}. Certain neurons, such as neuron 8, exhibited excessive responsiveness across all classes, while others remained underutilized, highlighting considerable redundancy and inefficiency in the network's behavior. 

Addressing this issue required increasing the network's complexity, which proved to be the key solution. The introduction of synaptic delays and a delay-learning mechanism allowed neurons to fine-tune their activation times based on incoming spikes. As a result, the network achieved more precise alignment with specific spatio-temporal patterns, ultimately fostering meaningful specialization. This development is thoroughly analyzed and presented in Sec.~\ref{sec:final_results}.

\begingroup

\renewcommand{\arraystretch}{1.5}
\begin{table}[h]
\centering
\small
\begin{tabular}{cccccccc}
\toprule
\textbf{$\langle{N_{\text{bg}}}\rangle$} & 
\textbf{$A_{-1}$ [\%]} & 
\textbf{$A_{+1}$ [\%]} & 
\textbf{$A_{-3}$ [\%]} & 
\textbf{$A_{+3}$ [\%]} & 
\textbf{$A_{-10}$ [\%]} & 
\textbf{$A_{+10}$ [\%]} & 
\textbf{$F$ [\%]} \\
\midrule
$50$  & $86 \pm 1$   & $76 \pm 1$   & $98.6 \pm 0.4$ & $99.9 \pm 0.2$ & $93.5 \pm 0.8$ & $96.4 \pm 0.7$ & $2.2 \pm 0.1$ \\
$100$ & $70 \pm 2$   & $61 \pm 2$   & $98.4 \pm 0.4$ & $98.9 \pm 0.4$ & $97.0 \pm 0.6$ & $97.6 \pm 0.5$ & $2.1 \pm 0.1$ \\
$200$ & $64 \pm 2$   & $49.4 \pm 2$ & $93.5 \pm 0.8$ & $92 \pm 1$     & $97.5 \pm 0.6$ & $97.4 \pm 0.6$ & $3.9 \pm 0.2$ \\
\bottomrule
\end{tabular}
\vspace{3mm}
\caption{\emph{Aggregate acceptance and fake rate of the networks for different patterns and average noise levels. The networks were trained and tested with the same average number of noise hits. Each test dataset contains $N_{\text{ev}}^{\text{test}} = 10,000$ events.}}
\label{tab:aggregate_results_thesis}
\end{table}
\endgroup

\FloatBarrier

\subsection{Hyperparameter Tuning and Genetic Algorithm} 
\label{sec:genetic_algorithm}
To improve the model's ability to recognize particle trajectories, we introduced synaptic delays, which required precise hyperparameter tuning. Due to the high dimensionality of the parameter space, we employed a genetic algorithm (GA) to efficiently optimize network configuration.
 
The optimization process focused on a set of key hyperparameters governing synaptic and neuronal dynamics. These include temporal scaling factors, reset potential adjustments, membrane potential thresholds, and learning rates for synaptic delay modulation. The genetic algorithm was designed to efficiently navigate this high-dimensional space by iteratively improving parameter sets based on performance criteria. Table~\ref{tab:hyperparameters_definition} summarizes the main hyperparameters considered in the optimization.

\begin{table}[h]
\centering
\small
\begin{tabular}{cc}
\toprule
\textbf{Parameter} & \textbf{Description} \\
\midrule
$K_\mu$, $\alpha$ & Temporal scaling factor and strength coefficient for IPSP dynamics. \\
$K_1$, $K_2$ & Scaling constants for the reset potential after neuron activation. \\
$T_0$, $T_1$ & Membrane potential thresholds for neuron activation in Layer 0 and Layer 1. \\
$d_{-}$, $d_{+}$ & Learning rates for synaptic delay depression and potentiation. \\
$\tau_m$, $\tau_s$ & Membrane potential decay time constant and synaptic potential time constant. \\
$\tau_{d_-}$, $\tau_{d_-}^{'}$ & Time constant and auxiliary time constant for synaptic delay depression. \\
$\tau_{d_+}$, $\tau_{d_+}^{'}$ & Time constant and auxiliary time constant for synaptic delay potentiation. \\
\bottomrule
\end{tabular}
\vspace{3mm}
\caption{\emph{Hyperparameters of the SNN, optimized for particle trajectory classification under high-noise conditions using a genetic algorithm.}}
\label{tab:hyperparameters_definition}
\end{table}
The optimization of these parameters is performed using NSGA-II (Non-dominated Sorting Genetic Algorithm II), a state-of-the-art multi-objective evolutionary algorithm~\cite{gad2021pygadintuitivegeneticalgorithm}. Implemented via the pyGAD library~\cite{996017}, NSGA-II is particularly suited for problems involving multiple conflicting objectives, such as maximizing network efficiency while minimizing the fake rate and maximizing selectivity. The algorithm employs a random mutation strategy and a single-point crossover mechanism to generate new candidate solutions, with parent selection based on non-dominated sorting to maintain population diversity across generations. Each iteration evaluates the fitness of candidate solutions by simulating the SNN and analyzing key performance metrics. The process continues until convergence towards an optimal set of hyperparameters is achieved, ensuring a robust and efficient network configuration.

\FloatBarrier
\subsection{Optimization Workflow for Spiking Neural Networks in Noisy Environments}
\label{sec:Optimization-Workflow}

To achieve a SNN capable of classifying particles with varying charges and momenta in dense and noisy environments, a phased optimization workflow was employed. This progressive strategy systematically exposes the network and the accompanying GA to increasingly complex tasks, ensuring robustness and generalization. The workflow is outlined as follows:

\begin{itemize}

    \item \textbf{Phase 1: Simplified Problem with Minimal Background}  
    \begin{itemize}
        \item Background noise is controlled, with the average number of background hits set to $\langle N_{\text{bg}}\rangle = 100$. 
        \item The GA optimizes the network for a reduced classification task focused on identifying negative muons with transverse momenta $p_T \in \{1, 3, 10\}\ \text{GeV}$, resulting in a three-class classification problem.
        \item Delay learning, a mechanism for modulating synaptic delays, is employed to enhance selectivity and minimize false positives.
        \item The hyperparameter search space for this phase includes all the ones defined in Table ~\ref{tab:hyperparameters_definition}.
    \end{itemize}

    \item \textbf{Phase 2: Incorporating Antimuon Classification}  

     \begin{itemize}
         \item The optimal network configuration from Phase 1 is selected as a baseline. The most active and specialized neurons are preserved, while their delay properties are mirrored and adjusted to initialize new neurons. Initial individuals in the GA are mutations of this configuration.
         \item Antimuon events are introduced into the training and validation datasets, expanding the classification task.
         \item The hyperparameter search space is refined by fixing the values of $K_1$, $K_2$, $\tau_m$, $\tau_s$, $\tau_{d_-}$, $\tau_{d_+}$, $\tau_{d_-}^{'}$, $\tau_{d_+}^{'}$, and constraining the ranges of $T_0$, $T_1$, $K_\mu$, $\alpha$.

         \item Delay learning rates ($d_\pm$) are reduced to allow finer convergence during training.         
     \end{itemize}
    
    \item \textbf{Phase 3: Evaluating Network Robustness Across Noise Levels}  
    
    \begin{itemize}
        \item The optimal network from Phase 2 undergoes additional refinement, where neurons with high activation and specialization are retained, and less significant neurons are pruned.
        \item The network is tested at varying noise levels to assess robustness. Based on these results, the network undergoes further retraining at elevated noise levels during the next phase.  
    \end{itemize}

    \item \textbf{Phase 4: Fine-tuning at High Noise Levels}  
    Results for this phase are presented in Sec.~\ref{sec:final_results}. 
    \begin{itemize}
         \item The optimized network from Phase 3 is selected as the basis for this phase. Initial individuals in the GA are mutations of this configuration.
         \item Background noise is increased, with the average number of background hits raised to $\langle N_{\text{bg}}\rangle = 300$.
         \item The hyperparameter search space is limited to $T_0$, $T_1$, $K_\mu$, $\alpha$, $\tau_m$, $\tau_s$, with narrower parameter ranges to focus on fine adjustments.
         \item The delay in learning rates ($d_\pm$) are further reduced to achieve incremental improvements in accuracy and robustness.
    \end{itemize}
\end{itemize}

\FloatBarrier
\section{Results}

\subsection{Single Track Pattern Recognition Under High Levels of Noise}
\label{sec:final_results}
\begin{figure}[h]
    \centering 
    \begin{subfigure}[b]{0.48\textwidth}
        \centering
        \includegraphics[width=\linewidth]{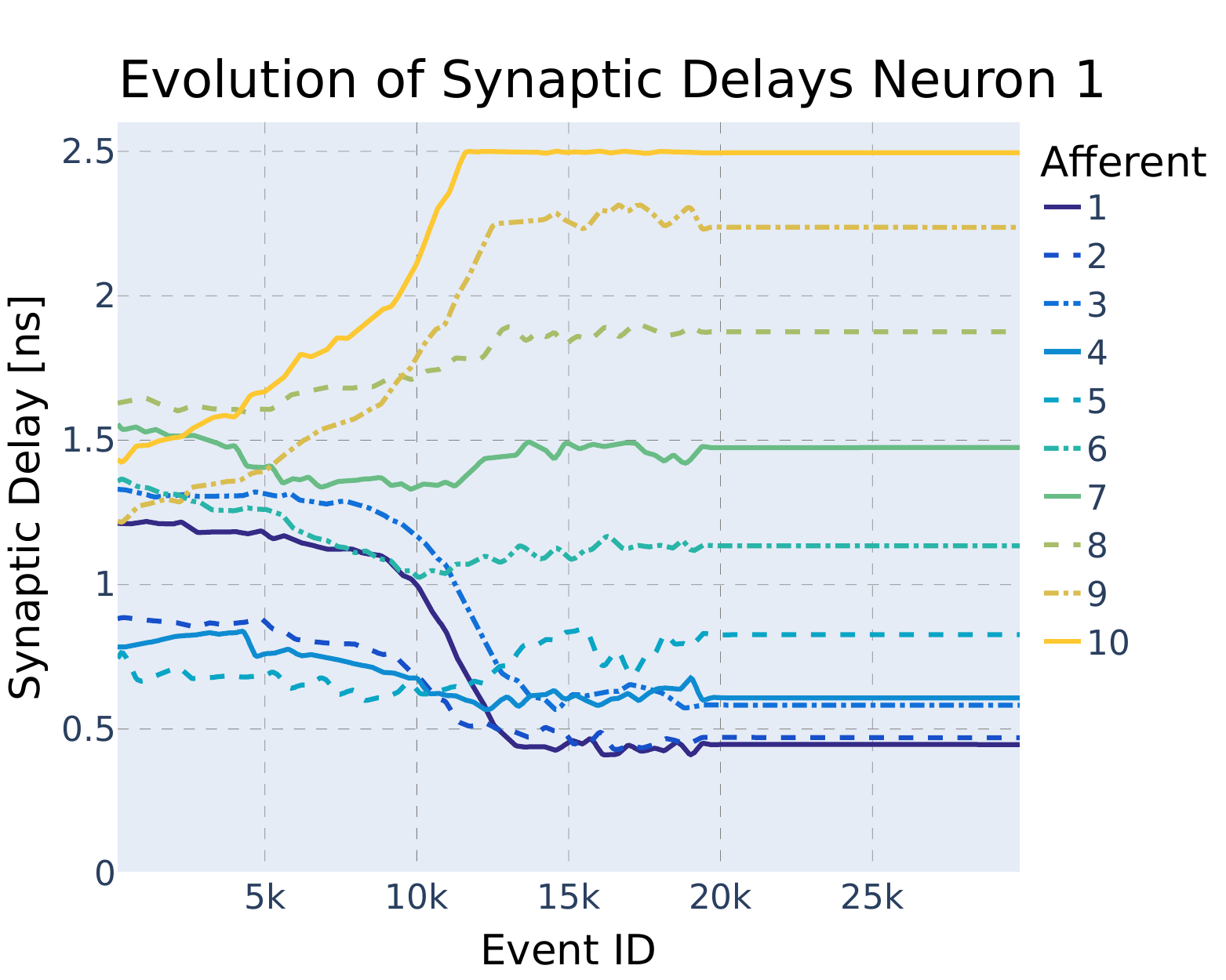}
        \caption{}
        \label{fig:delay-evolution-3}
    \end{subfigure}
    \hfill
    \begin{subfigure}[b]{0.48\textwidth}
        \centering
        \includegraphics[width=\linewidth]{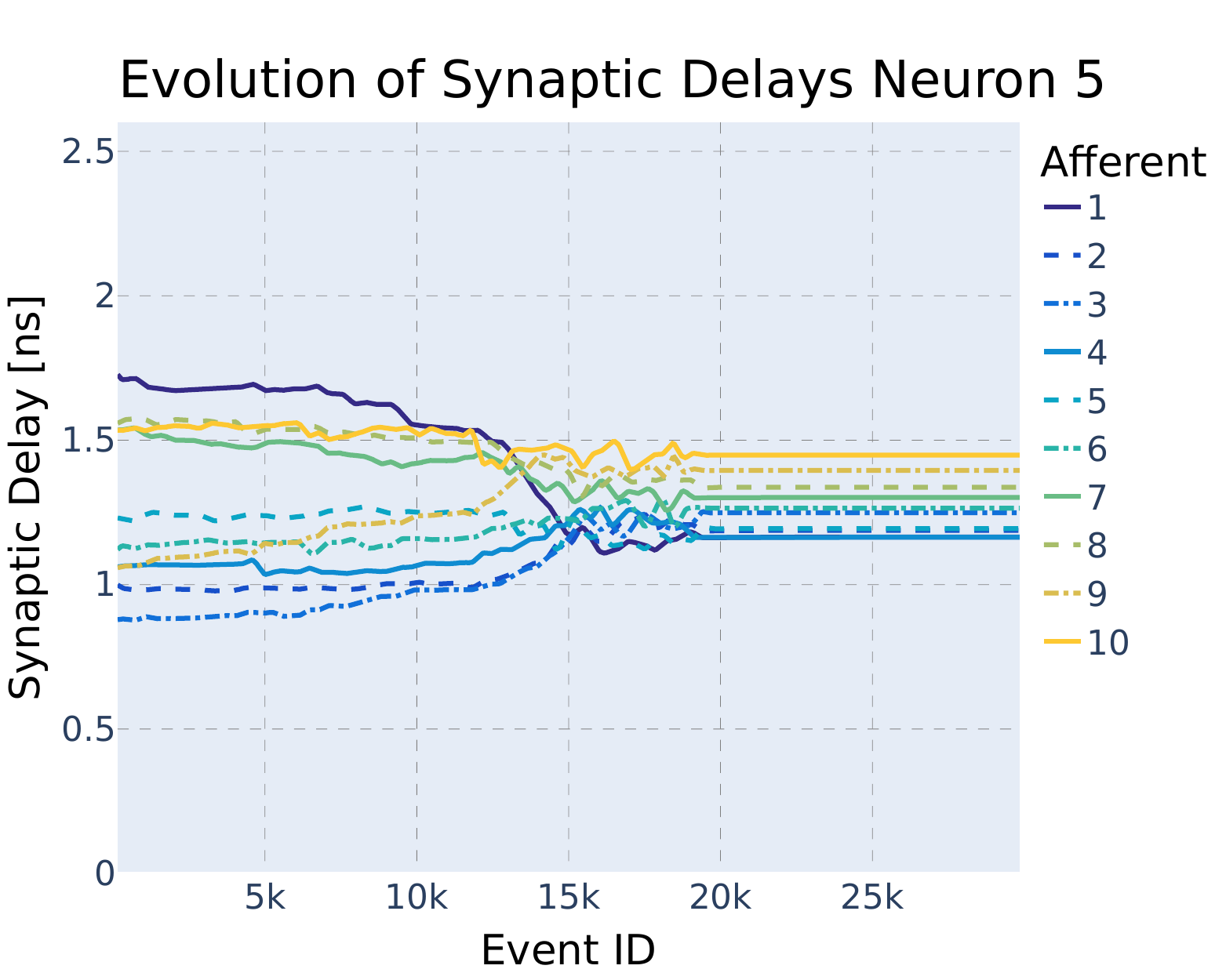}
        \caption{}
        \label{fig:delay-evolution-1}
    \end{subfigure}
    \caption{\emph{Evolution of synaptic delays over the course of training for neurons specializing in recognizing particles with transverse momentum values of 1 GeV (a),  and 10 GeV (b). The x-axis represents the training iteration, and the y-axis shows the delay value for each afferent.}}
    \label{fig:delay-evolution-summary}
\end{figure}

\begin{figure}
    \centering
    \begin{subfigure}[t]{0.48\linewidth}
        \centering
        \includegraphics[width=\linewidth]{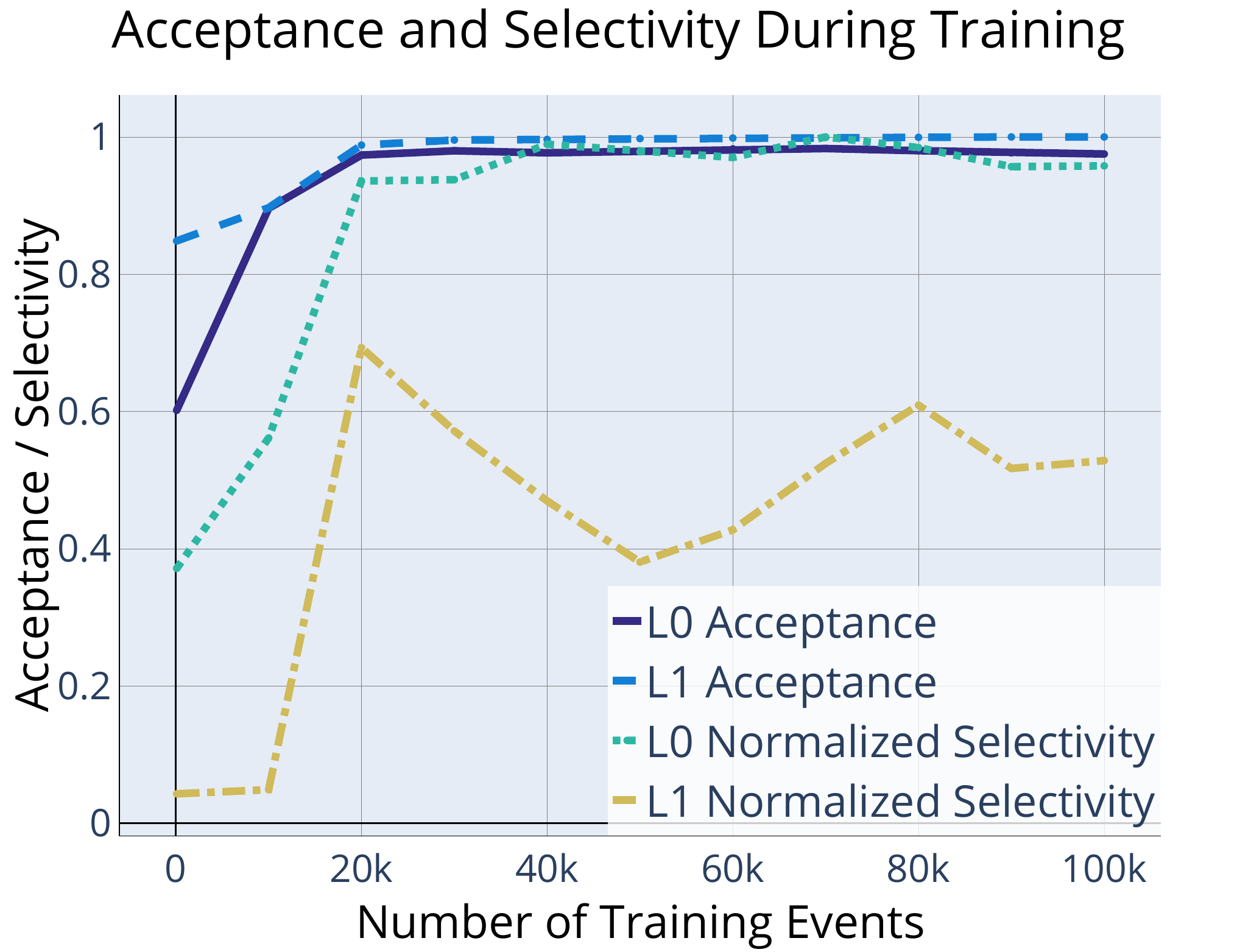}
        \caption{\emph{Evolution of acceptance and selectivity of the SNN during the learning process. The X-axis represents the number of events processed, whereas the Y-axis denotes the acceptances and selectivities of L0, L1 neurons. The selectivities have been normalized by their maximum values for visualization purposes.}}
        \label{fig:efficiency_fake_rate}
    \end{subfigure}
    \hfill
    \begin{subfigure}[t]{0.48\linewidth}
        \centering
        \includegraphics[width=\linewidth]{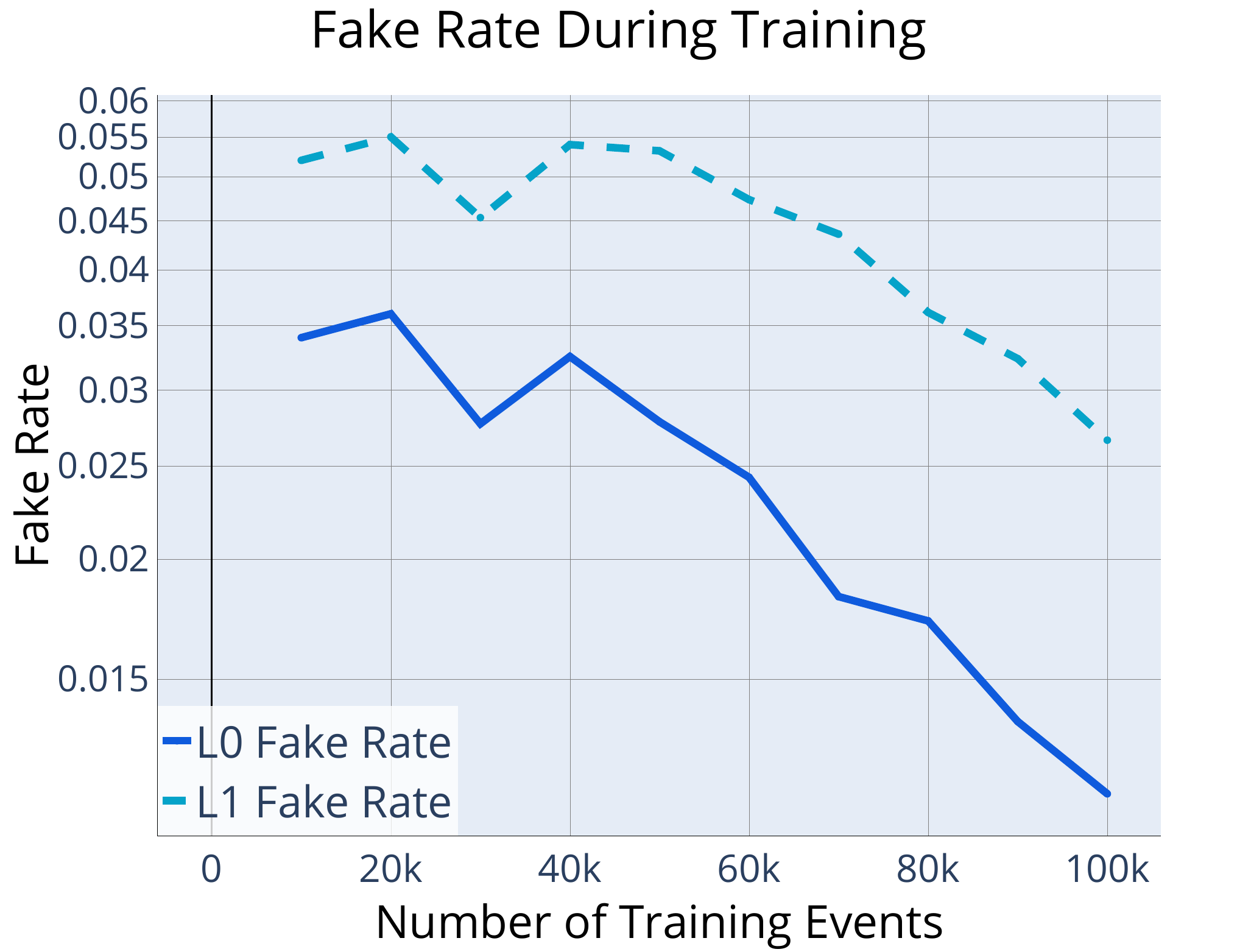}
        \caption{\emph{Evolution of the fake rates of neurons during the learning process. The X-axis represents the number of events processed, whereas the Y-axis denotes the fake rates of L0, L1 neurons.}}
        \label{fig:fake_evolution}
    \end{subfigure}
    
    \caption{\emph{Comparison of acceptance, selectivity, and fake rate evolution during the learning process.}}
    \label{fig:efficiency_selectivity_comparison}
\end{figure}

\begin{figure}[h]
    \centering
    \begin{subfigure}[b]{0.48\linewidth}
        \centering
        \includegraphics[width=\linewidth]{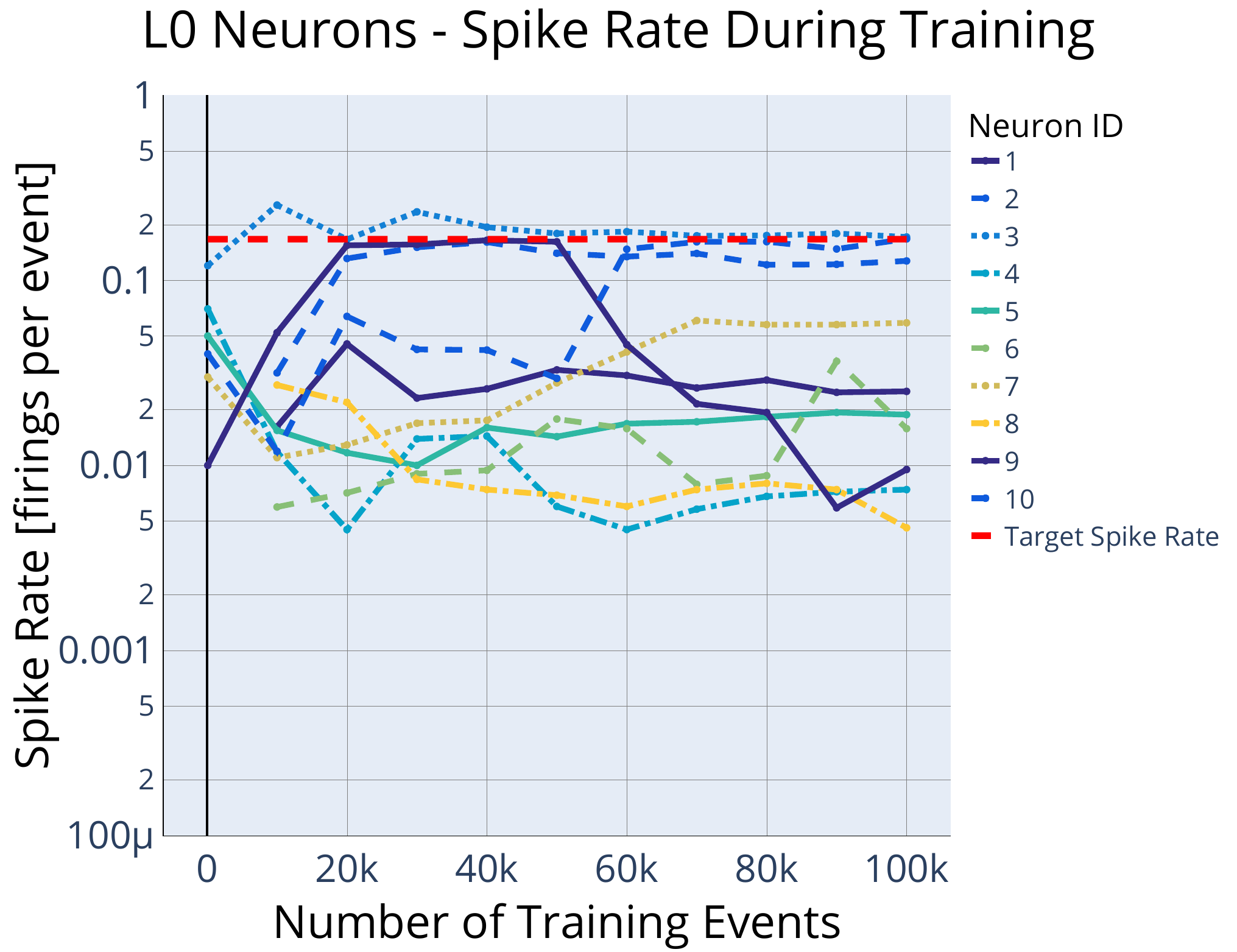}
        \caption{}
        \label{fig:L0 firing rate}
    \end{subfigure}
    \begin{subfigure}[b]{0.48\linewidth}
        \centering
        \includegraphics[width=\linewidth]{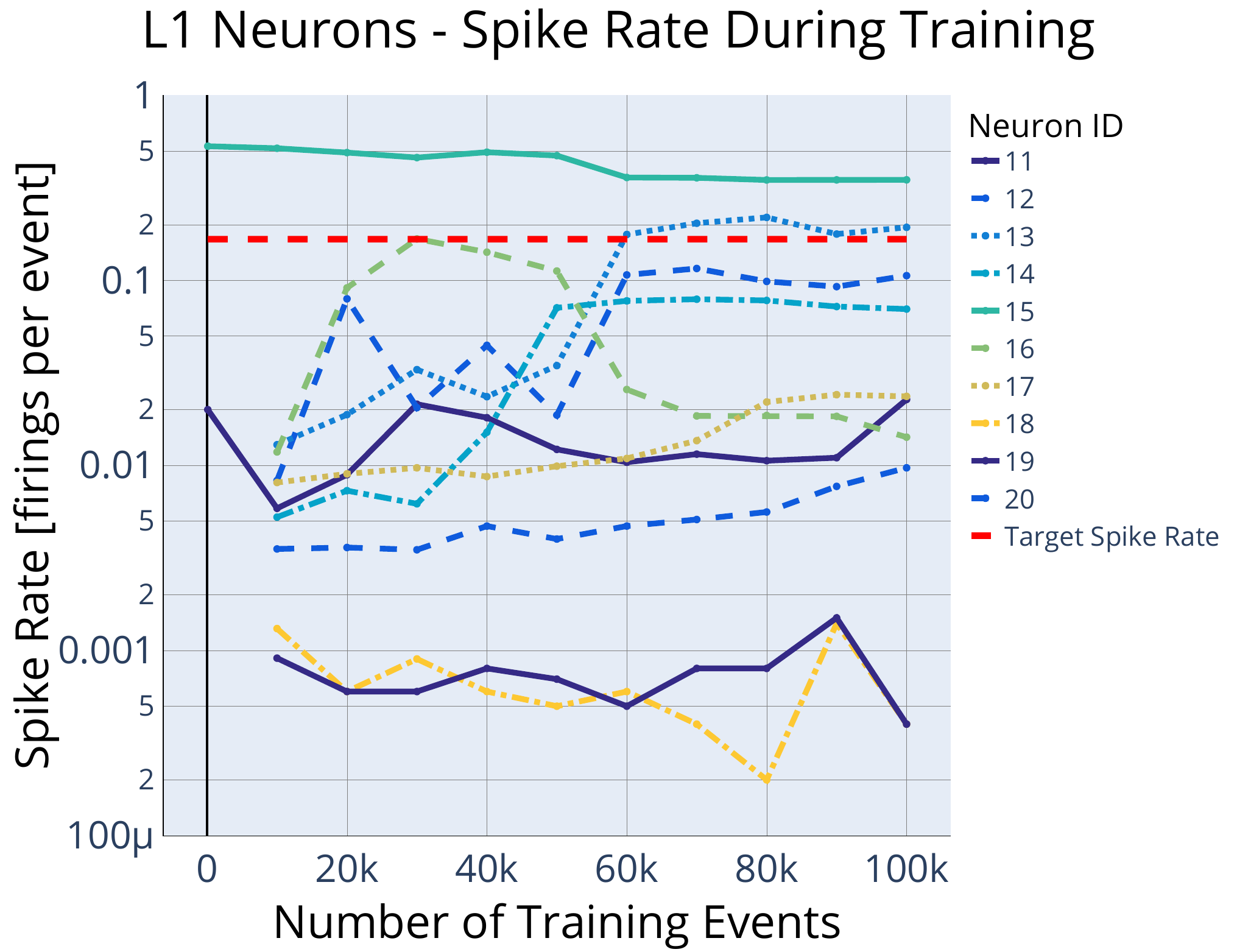}
        \caption{}
        \label{fig:L1 firing rate}
    \end{subfigure}
    \caption{\emph{Comparison of spike rates, defined as the number of neuron firings per event, throughout the learning process. The X-axis represents the number of processed events. The left Y-axis shows the firing rate of L0 neurons, while the right Y-axis displays the firing rate of L1 neurons on a logarithmic scale. The target spike rate represents the expected value for a perfectly specialized neuron and corresponds to $r = \frac{1}{6} \frac{\text{firings}}{\text{event}}$.}}
    \label{fig:firing_rate_comparison}
\end{figure}

Before presenting the final results under high noise conditions, we first highlight key findings of Phases 1 and 3, which shaped our approach to improving the optimization process.
Figure~\ref{fig:delay-evolution-summary} illustrates the evolution of the synaptic delays during Phase 1 for three neurons specializing in the recognition of muons with distinct transverse momenta ($p_T = 1$, $3$, and $10\ \text{GeV}$). By processing a total of 20,000 training events, the delays transitioned from random initial values to stable configurations in an unsupervised manner, effectively aligning the network with the corresponding spatio-temporal patterns.
Figures~\ref{fig:efficiency_selectivity_comparison} illustrate the evolution of acceptance, selectivity, and fake rate as functions of the number of training events processed by an SNN initialized with the same hyperparameters as the best-performing individual from Phase 1. The plots indicate that after 20,000 training events, both acceptance and selectivity are already near their maximum values, while extended training could further reduce fake rates exponentially. This trend is further supported by Fig.~\ref{fig:L0 firing rate}, which demonstrates how three L0 neurons progressively adjust their firing rates toward the expected value for a perfectly specialized neuron under these conditions (red dashed line), while the other firing rates are lower by an order of magnitude. Notably, these three neurons correspond to the most specialized units within the analyzed network.


\begin{figure}[h]
    \centering
    \begin{subfigure}[b]{0.48\textwidth}
        \centering
        \includegraphics[width=\linewidth]{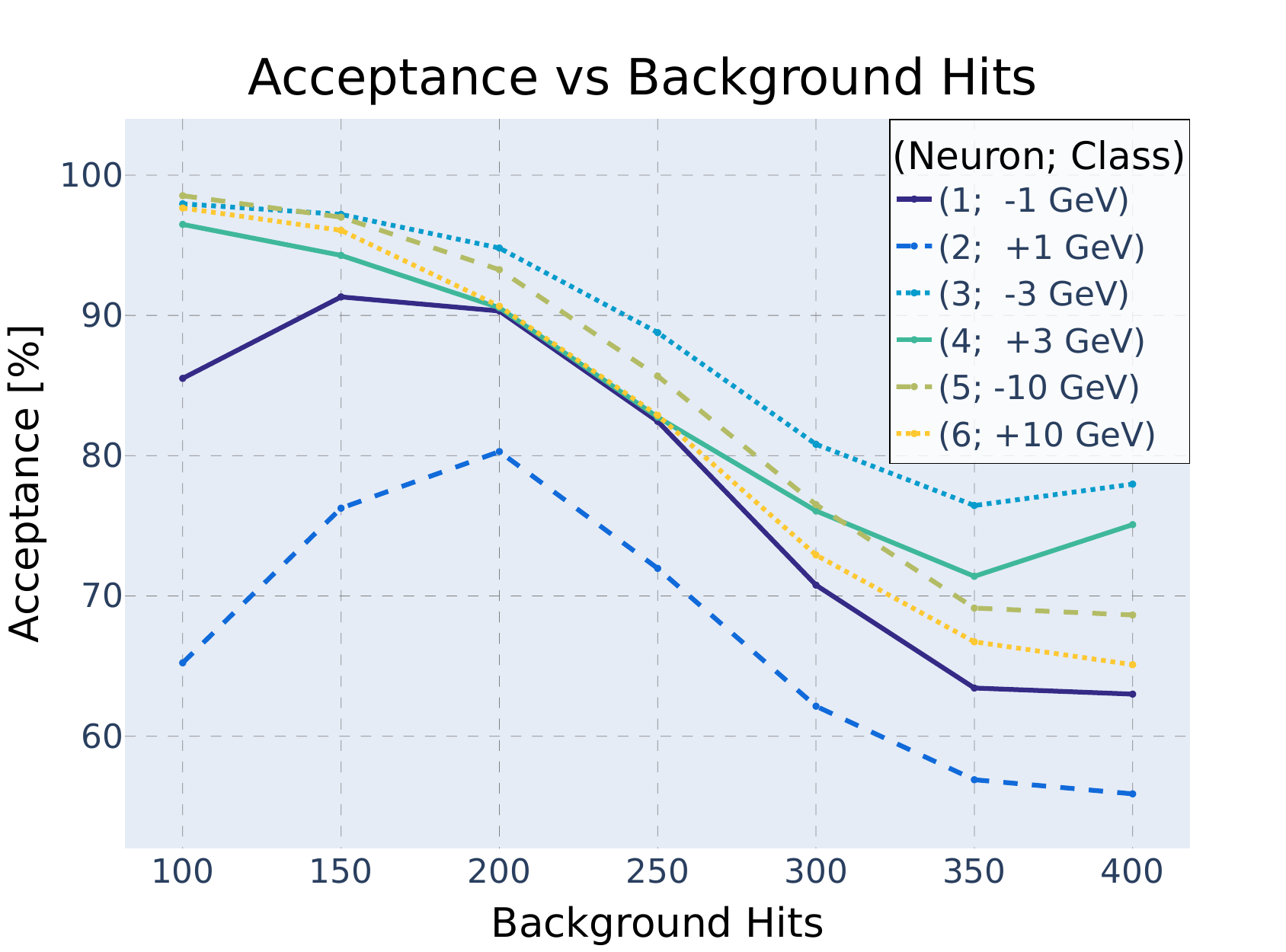}
        \caption{\emph{Acceptance scaling with background level.}}
        \label{fig:efficiency_bg-hits}
    \end{subfigure}
    \hfill
    \begin{subfigure}[b]{0.48\textwidth}
        \centering
        \includegraphics[width=\linewidth]{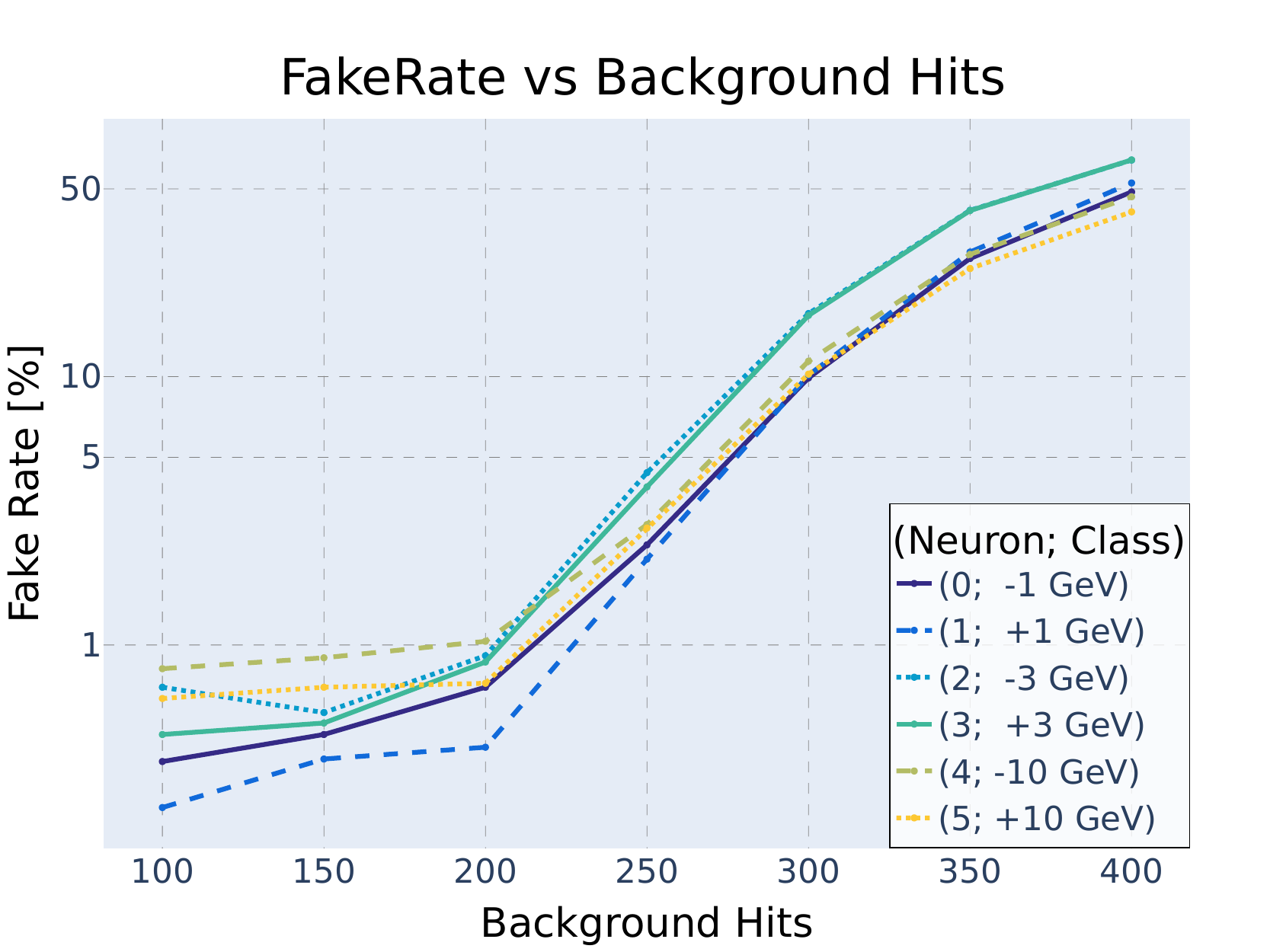}
        \caption{\emph{Fake rate scaling with background level.}}
        \label{fig:FakeRate_bg-hits}
    \end{subfigure}
    \caption{\emph{Performance of a network trained with a background level of $\langle N_{\text{bg}}\rangle = 100$ when tested at higher background levels.}}
    \label{fig:scaling_behavior}
\end{figure}

In Phase 3, the robustness of the network was tested by training it on events with $\langle N_{\text{bg}} \rangle = 100$ noise hits and then evaluating its performance under varying noise levels. The outcomes, summarized in Fig.~\ref{fig:scaling_behavior}, reveal a general decline in neuron acceptances and an exponential increase in fake rates for $\langle N_{\text{bg}}^{\text{test}} \rangle \geq 200$ noise hits. These results underscore that network performance is closely related to noise levels in the training environment. In particular, specific time constants that regulate the build-up of neuron potentials such as $\tau_m$ and $\tau_s$ are crucial in determining the robustness of the network at different levels of noise.

\FloatBarrier
For the reason mentioned above, the SNN was re-optimized and assessed under high-noise conditions by simulating an average of 300 background clusters per event ($\langle N_{\text{bg}}\rangle = 300$). This noise level represents a more challenging environment, and a more realistic one for the Phase-2 running conditions of the LHC. In fact, studies conducted on the central region of the CMS Phase-1 detector for the 
$t\Bar{t}$ process at a pileup level of 60 estimate that each signal particle generates, on average, one cluster for every eight clusters originating from noise~\cite{cms:TDR-014}. By extending these findings to the Barrel geometry of the Phase-2 tracker, the noise contribution per particle is extrapolated to approximately 80 clusters. Consequently, simulating 300 background clusters per event provides a meaningful and interesting test scenario for evaluating the network's capacity to identify true particle trajectories amidst significant combinatorial noise. 
Figure~\ref{fig:signal-300} provides an example of a transverse plane projection of an event with this level of noise. A complete list of the final hyperparameters is available in Table~\ref{tab:hyperparameters}.
The aggregate performance of the network in terms of signal acceptance (\(A\)), fake rate (\(F\)) and selectivity (\(S\)) is summarized in Table~\ref{tab:aggregate_results_300br}. Key findings include:
\begin{itemize}
    \item \textbf{High acceptance:} The SNN achieved greater than 98\% acceptance for all transverse momentum classes, with peak values reaching 100\% for particles with $p_T = 10\ \text{GeV}$.
    \item \textbf{Low fake rate:} The network effectively suppressed noise, achieving a fake rate of approximately \(3.0 \pm 0.2\%\).
    \item \textbf{Strong selectivity:} A selectivity score of 3.48 indicates robust discrimination between signal and noise patterns.
\end{itemize}

\vspace{-0.7cm}
\begingroup
\renewcommand{\arraystretch}{1.5}
\begin{table}[h!]
\centering
\resizebox{\textwidth}{!}{%
\begin{tabular}{ccccccccc}
\hline
\textit{$\boldsymbol{\langle{N_{\text{bg}}^{\text{test}}}\rangle}$} &
\textit{$\boldsymbol{A_{-1}}$} [\%] &
\textit{$\boldsymbol{A_{+1}}$} [\%] & 
\textit{$\boldsymbol{A_{-3}}$} [\%] &
\textit{$\boldsymbol{A_{+3}}$} [\%] &
\textit{$\boldsymbol{A_{-10}}$} [\%] &
\textit{$\boldsymbol{A_{+10}}$} [\%] &
\textit{$\boldsymbol{F}$} [\%] &
\textit{$\boldsymbol{S}$}
\\
\hline
\textit{$300$} & \textit{$98.2\pm0.3$} & \textit{$98.2\pm0.3$} & \textit{$99.90\pm0.08$} & \textit{$99.2\pm0.2$} & \textit{$100.00\pm0.05$} &
\textit{$100.00\pm0.05$} &
\textit{$3.0\pm0.2$} &
\textit{$3.48$} \\
\hline
\end{tabular}}
\vspace{2mm}
\caption{\emph{Aggregate Acceptance and Fake Rate of the network for different patterns and average noise level. Each test dataset contains $N_{\text{ev}}^{\text{test}} = 25,000$ events.}}
\label{tab:aggregate_results_300br}
\end{table}
\endgroup
\vspace{-1cm}

\begin{figure}[ht]
    \centering
    \includegraphics[width=0.8\linewidth]{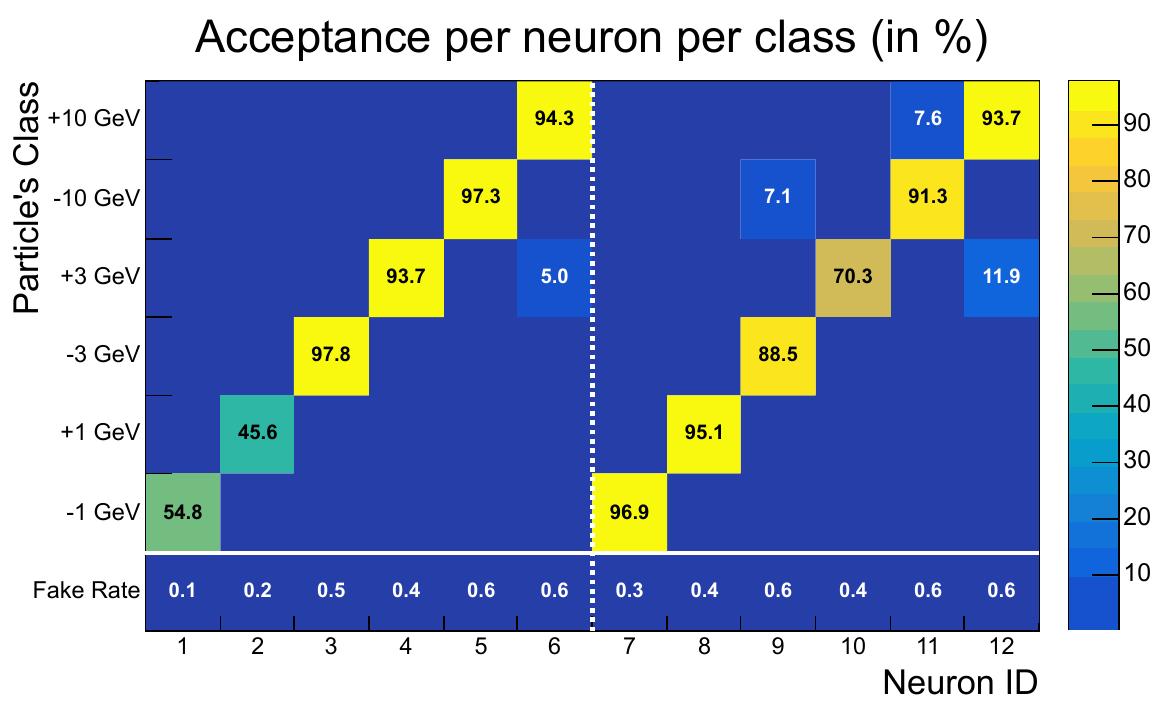}
    \caption{\emph{Heatmap of neuron activations across particle classes, with neurons represented on the x-axis and particle classes (characterized by charge and transverse momentum) on the y-axis. The color intensity indicates the acceptance rate, or the fraction of events in which the neuron was activated for a given class. The fake rate (false positives for noise-only events) is shown in the bottom row. This visualization demonstrates the network's ability to specialize neurons for specific classes while maintaining low false-positive rates.}}
    \label{fig:efficiencies_300br}
\end{figure}

Figure~\ref{fig:efficiencies_300br} presents the final acceptance of individual neurons across the different particle classes. The results demonstrate the remarkable specialization of neurons in recognizing specific patterns associated with distinct particle trajectories, even under high-noise conditions. The analysis highlights the performance of the SNN in classifying particle trajectories, even in high-noise conditions. Each neuron exhibits a distinct preference for a specific particle class, with at least one neuron achieving an acceptance rate exceeding 90$\%$ for every class. This specialization indicates that the network effectively learns to recognize characteristic spatiotemporal patterns associated with different transverse momentum $p_T$  values and charge states. Furthermore, the network demonstrates a low level of confusion across classes, with overlap in neuronal responses remaining below 5$\%$ in most cases. This ability to maintain clear distinctions between particle classes underscores the network’s effectiveness in accurately discriminating between different trajectories. Additionally, the fake rate per neuron, as illustrated in the first row of the heatmap, is consistently low, on the order of $10^{-3}$ , which highlights the robustness of the network in suppressing background noise even under challenging conditions.

\begin{figure}[!h]
    \centering
    \includegraphics[width=0.8\linewidth]{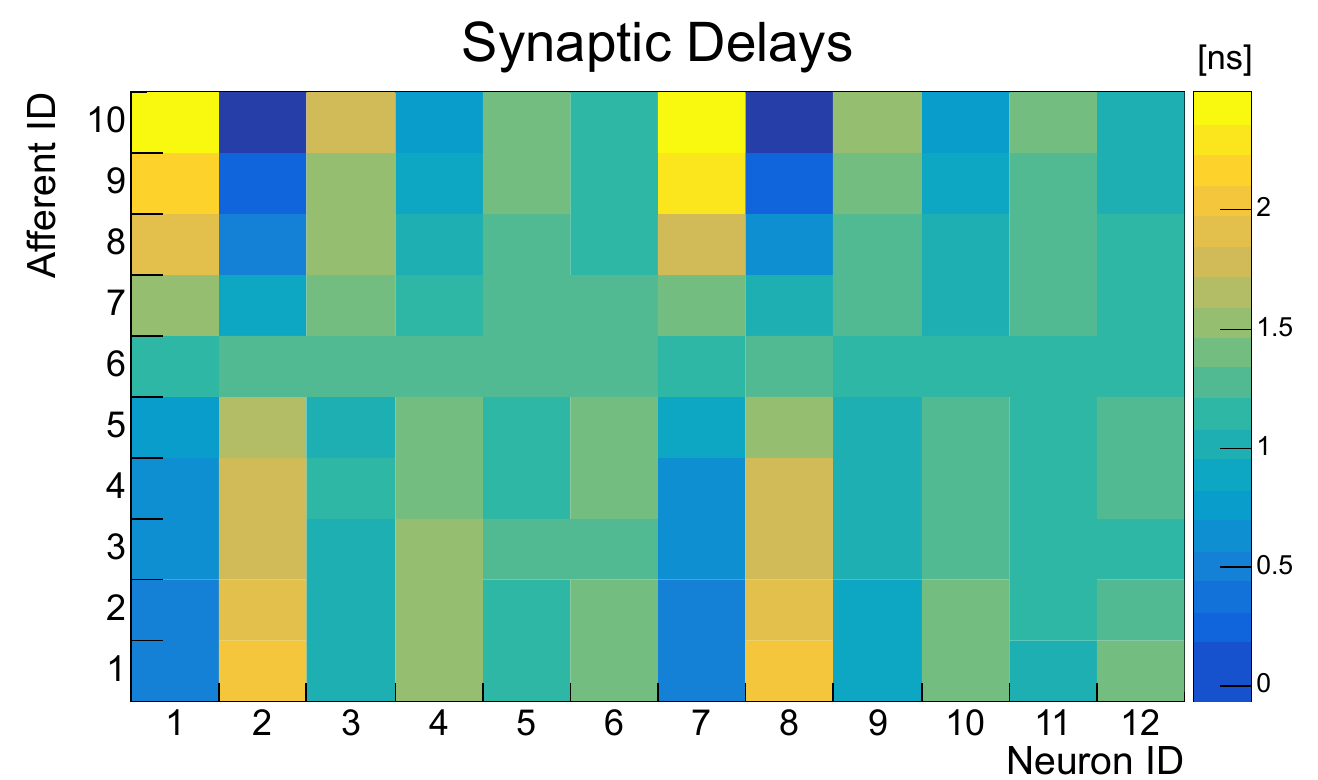}
    \caption{\emph{Synaptic delays of the different neurons. The x-axis represents the neuron ID, and the y-axis corresponds to the afferent ID. The color intensity indicates the synaptic delay (in nanoseconds), with brighter colors representing longer delays. This figure showcases the specialization of synaptic delays for different neurons to detect distinct spatiotemporal patterns in the input. }}
    
    \label{fig:delays}
\end{figure}

\begin{figure}[H]
    \centering
    \includegraphics[width=0.8\textwidth]{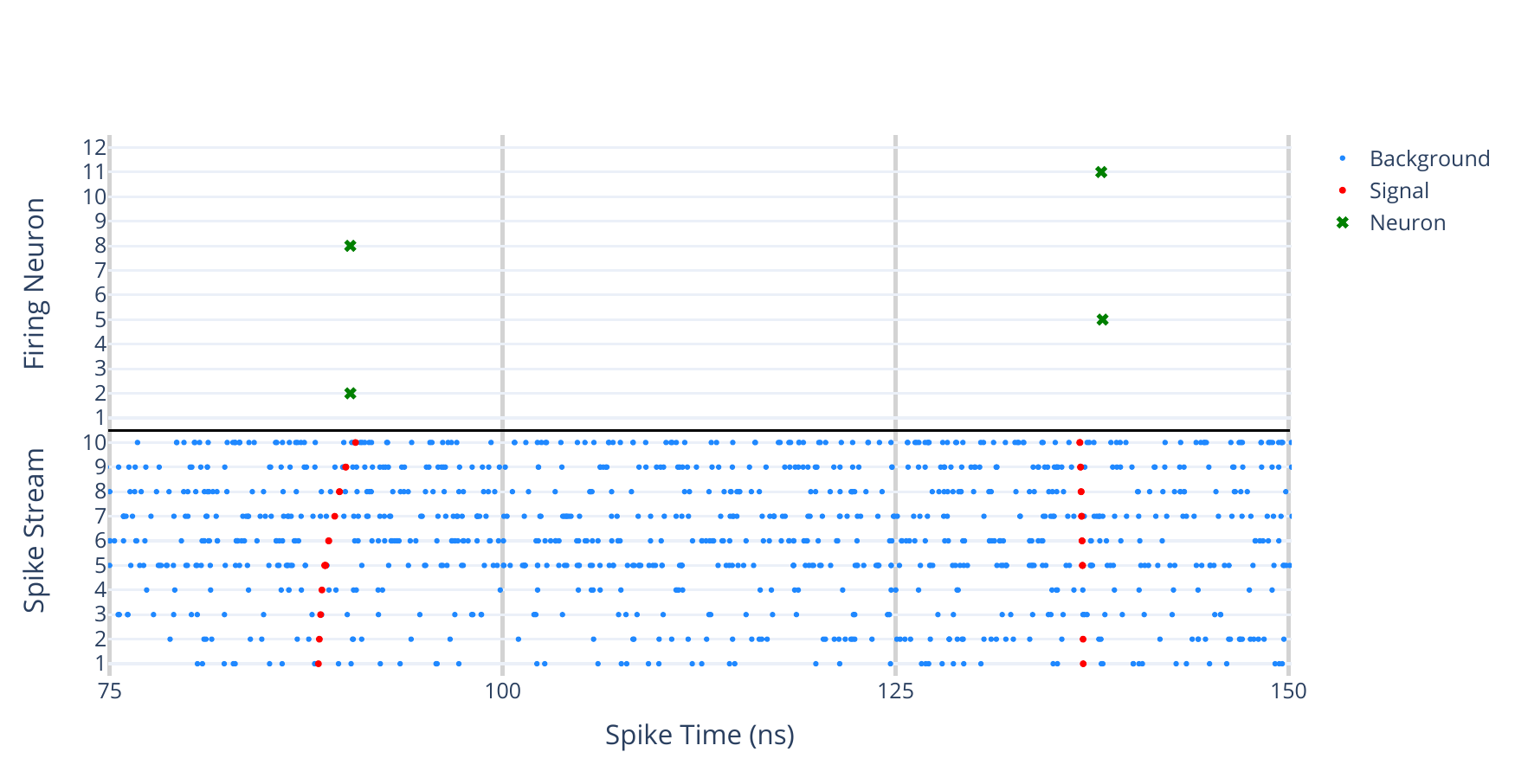} 
    \caption{\emph{Spatiotemporal spike patterns. In the bottom graph, the x-axis represents encoding time, while the y-axis designates each afferent. Each dot symbolizes an incoming spike from the associated afferent. Blue dots denote spikes associated with clusters of background, whereas red dots represent spikes associated with clusters of signal. In the top graph, the x-axis represents encoding time, while the y-axis designates each neuron. The green dots represent the activation of a neuron.}}
    \label{fig:neuron_response}
\end{figure}

The network's reliance on synaptic delays to temporally align spike-based inputs with neuronal firing played a critical role in achieving these results. Figure~\ref{fig:delays} illustrates the final synaptic delay configurations between neurons, revealing distinct temporal specializations for different classes of particles. In particular, neurons that specialize in lower $p_T$ patterns exhibit a wider spread in the magnitudes of synaptic delays compared to those that specialize in higher $p_T$ patterns. 
Figure~\ref{fig:neuron_response} illustrates the spatiotemporal spike patterns under high noise conditions and the activation of different neurons within the network. It is particularly noticeable that neurons activate in a time window close to the arrival of spikes associated to signal hits, whereas they do not activate in periods primarily occupied by spikes associated to noise.

While the single-track results demonstrate the feasibility of neuromorphic computing for real-time track reconstruction, high-energy physics experiments require the reconstruction of multiple overlapping particle trajectories per event. The next section proposes a preliminary study on how the proposed SNN model scales to multi-track events.

\clearpage

\subsection { Searching for Multiple Tracks } 
\begin{figure}[!h]
    \centering
    \includegraphics[width=0.8\linewidth]{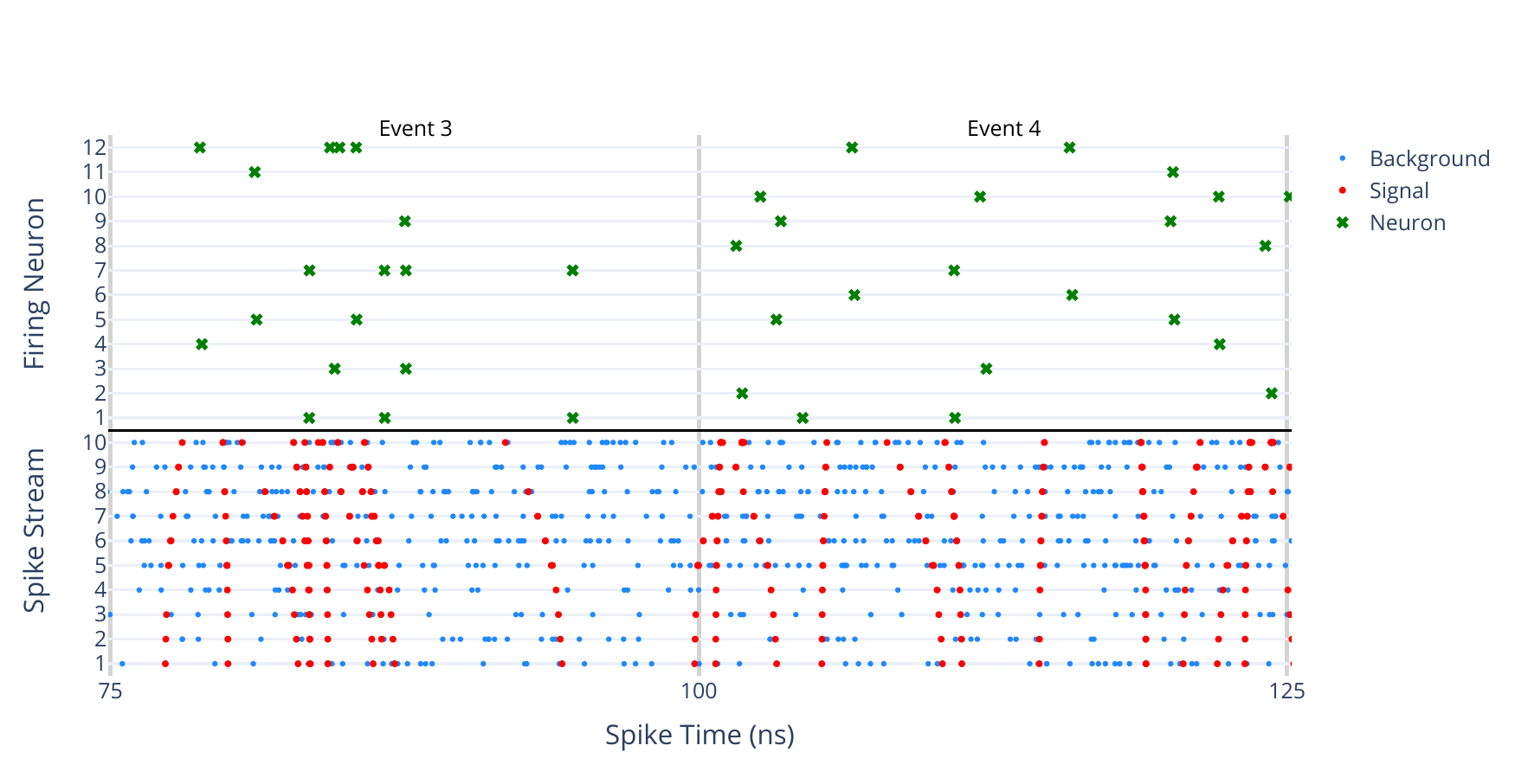}
    \caption{\emph{Spatio-temporal spike patterns in events containing 10 tracks. In the bottom graph, the x-axis represents encoding time, while the y-axis designates each afferent. Each dot symbolizes an incoming spike from the associated afferent. Blue dots denote spikes associated with clusters of background, whereas red dots represent spikes associated with clusters of signal. In the top graph, the x-axis represents encoding time, while the y-axis designates each neuron. The green dots represent the activation of a neuron.}}
    \label{fig:double-tracks-streams}
\end{figure}

Although the network described in Sec.~\ref{sec:final_results} was specifically trained for the reconstruction of single tracks, no fundamental limitations were identified that would prevent its extension to multiparticle tracking within the same event. Figure~\ref{fig:double-tracks-streams} presents an example of spatio-temporal spike patterns for events containing 10 tracks, demonstrating that, even in this more complex scenario, neurons activate according to the input tracks. This behavior arises from the fact that the SNN processes events as a time series, primarily relying on local information near the tracks. As a result, the primary factor affecting performance is the angular separation between tracks. When two tracks are too close in $\Phi$, the following challenges may arise:

\begin{itemize} 
    \item \textbf{Temporal Overlap:} Tracks appearing within the same time window may produce closely spaced spikes, making it difficult for the network to distinguish separate trajectories.

    \item \textbf{Refractory Periods:} Once a neuron fires in response to one track, it enters a refractory period, which may hinder the detection of other nearby tracks.  
    
    \item \textbf{Neuron Competition:} Lateral inhibition is a crucial mechanism that prevents multiple neurons from firing for the same pattern, allowing them to specialize. However, when two patterns occur within the same time window, this mechanism can become counterproductive, reducing the network’s ability to specialize effectively.  
\end{itemize}

In order to analyze this relationship, we examined eight datasets that were considered for the study, each of which contained 25,000 double-track events, with a fixed angular separation between them, defined as the difference of the average azimuthal angle of the two tracks. Furthermore, we define the following metrics. 

\begin{definition}[Detection Efficiency]
    Detection efficiency is the ratio of correct activations ({\it i.e.}, events when a neuron fires following the signal of a track of the class it was assigned to) to the total number of events of that class in the dataset:
    \begin{equation*}
        \epsilon = \frac{TP}{TP + FN}
    \end{equation*}
    where:
    \begin{itemize}
        \item $TP$ (True Positives) are correctly classified events.
        \item $FN$ (False Negatives) are missed events.
    \end{itemize}
\end{definition}

\begin{figure}
    \centering
    \begin{subfigure}[t]{0.45\linewidth}
        \centering
        \includegraphics[width=\linewidth]{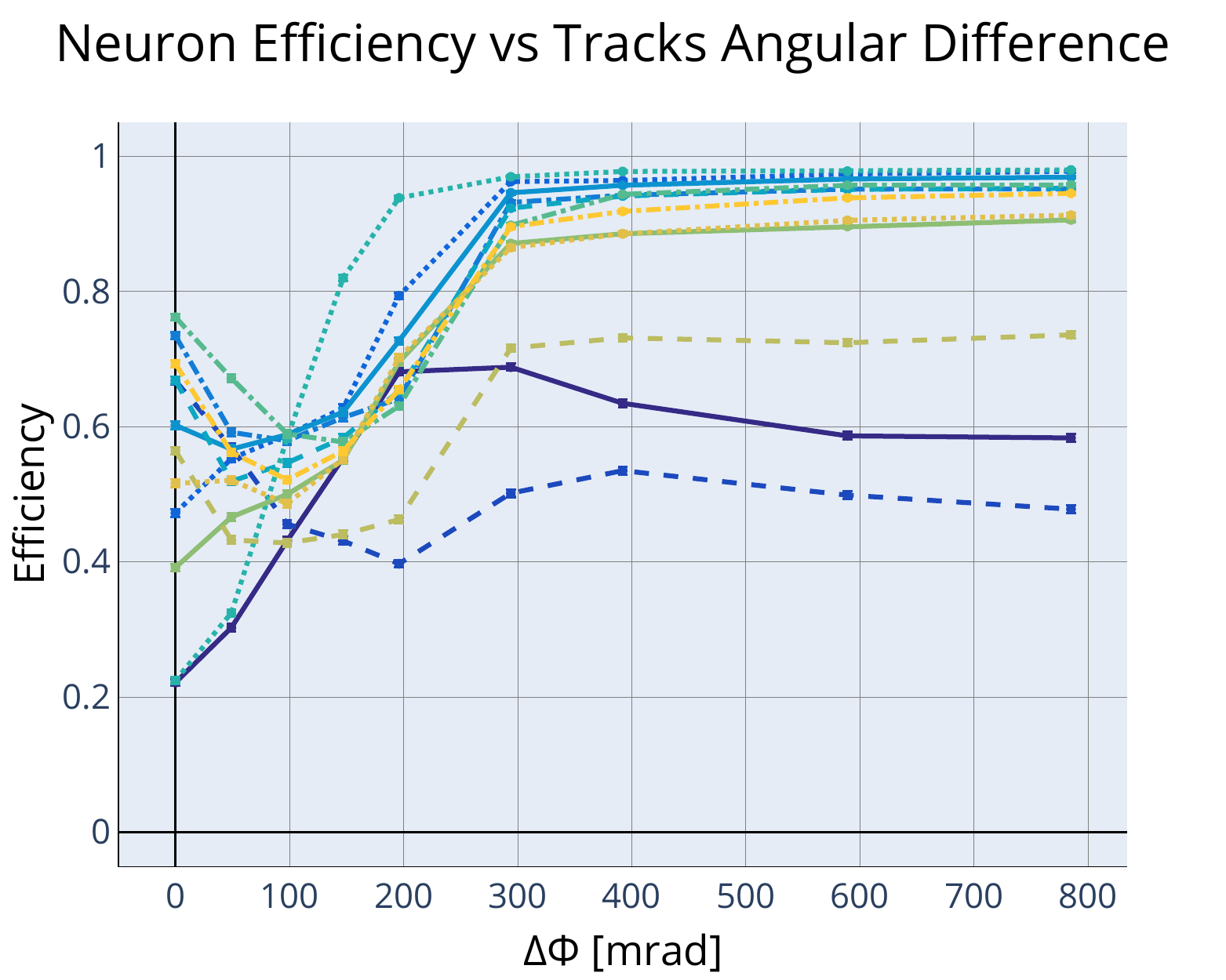}
        \caption{\emph{Detection efficiency of each neuron.}}
        \label{fig:neuron_efficiency_vs_angle}
    \end{subfigure}
    \hfill
    \begin{subfigure}[t]{0.45\linewidth}
        \centering
        \includegraphics[width=\linewidth]{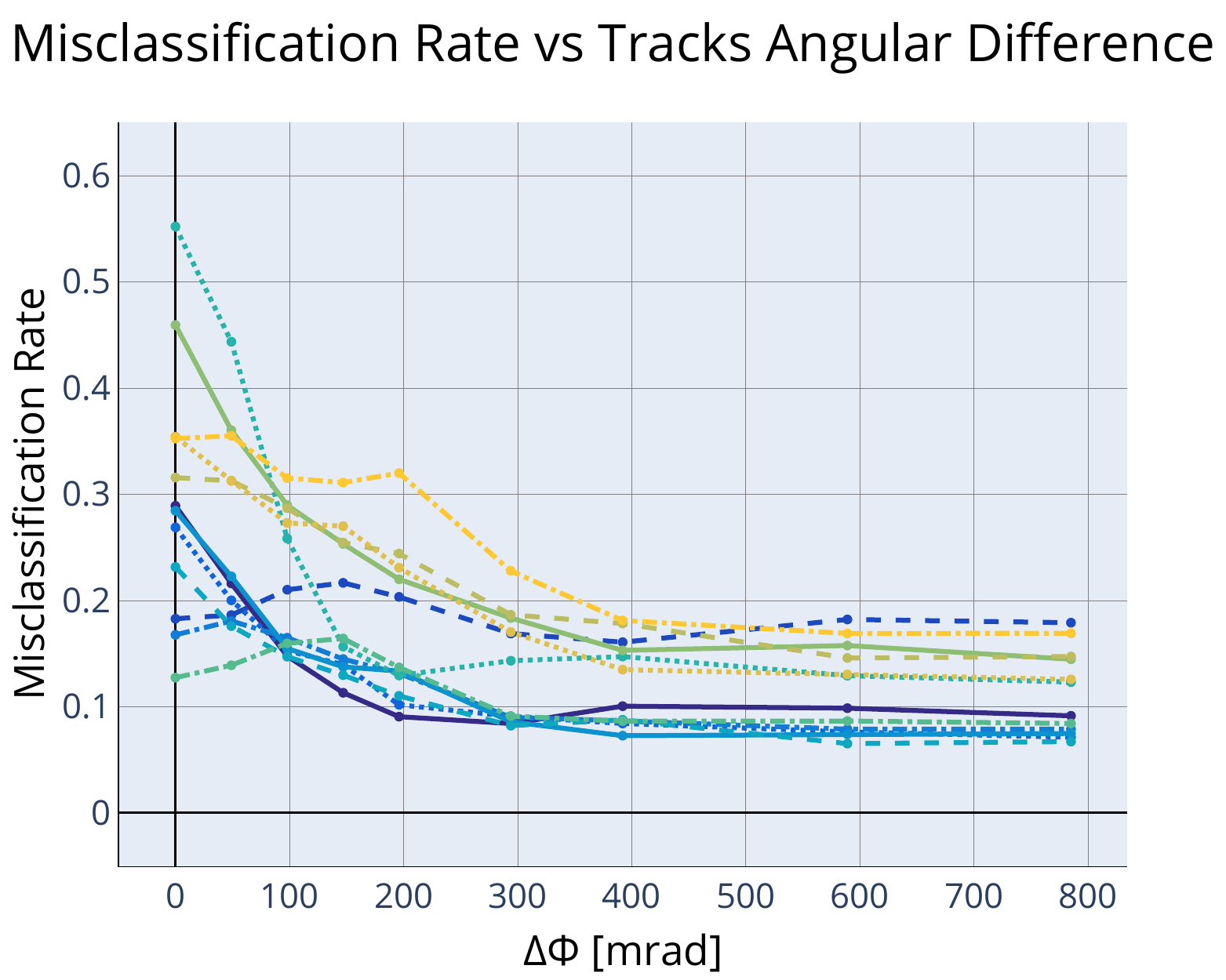}
        \caption{\emph{Misclassification rate of each neuron.}}
        \label{fig:misclassification_rate_vs_angle}
    \end{subfigure}
    
    \vspace{0.5em} 
    
    \begin{subfigure}[t]{0.45\linewidth}
        \centering
        \includegraphics[width=\linewidth]{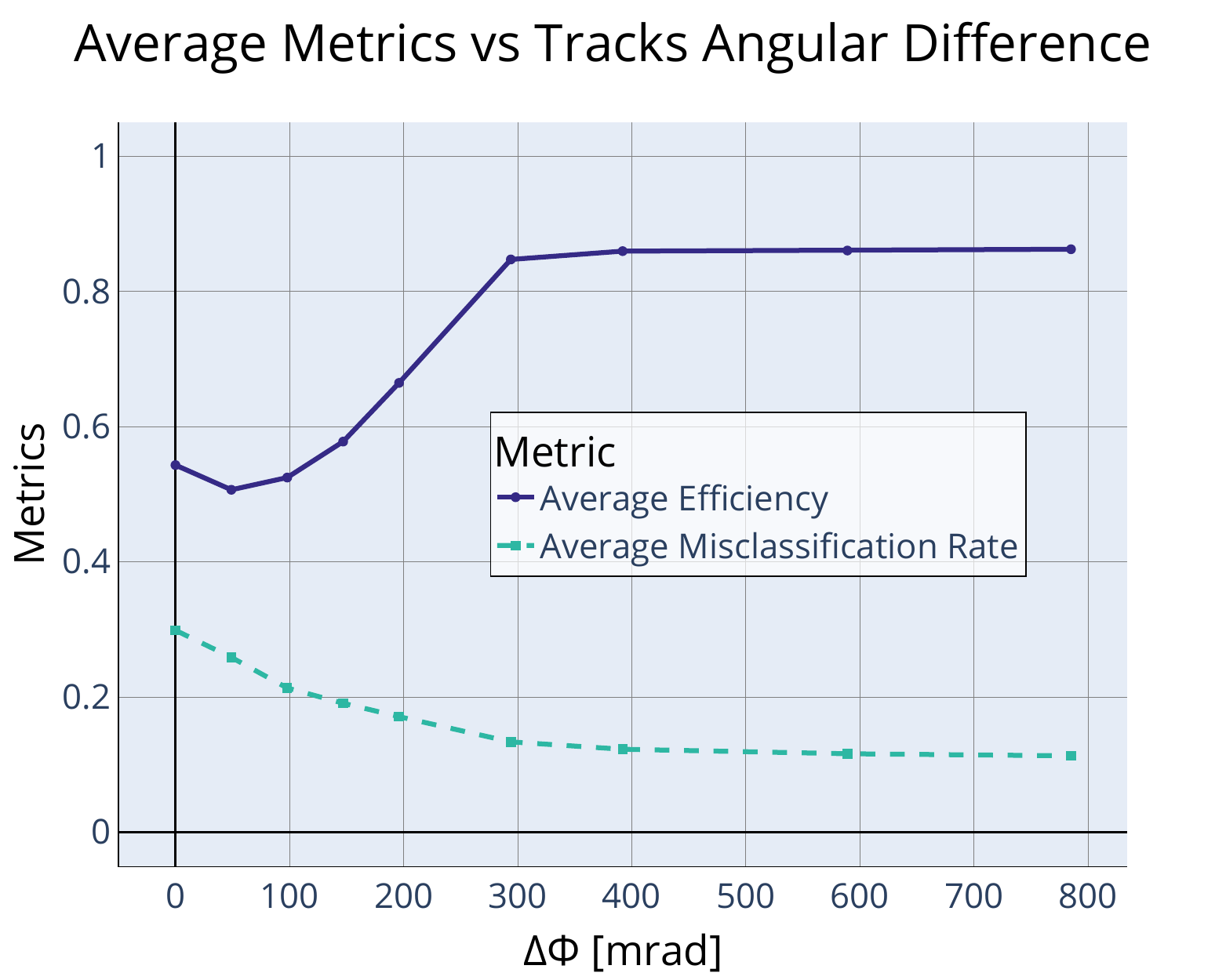}
        \caption{\emph{Average misclassification rate and detection efficiency of the SNN.}}
        \label{fig:angulare_difference_study}
    \end{subfigure}
    \hfill
    \begin{subfigure}[t]{0.45\linewidth}
        \centering
        \includegraphics[width=\linewidth]{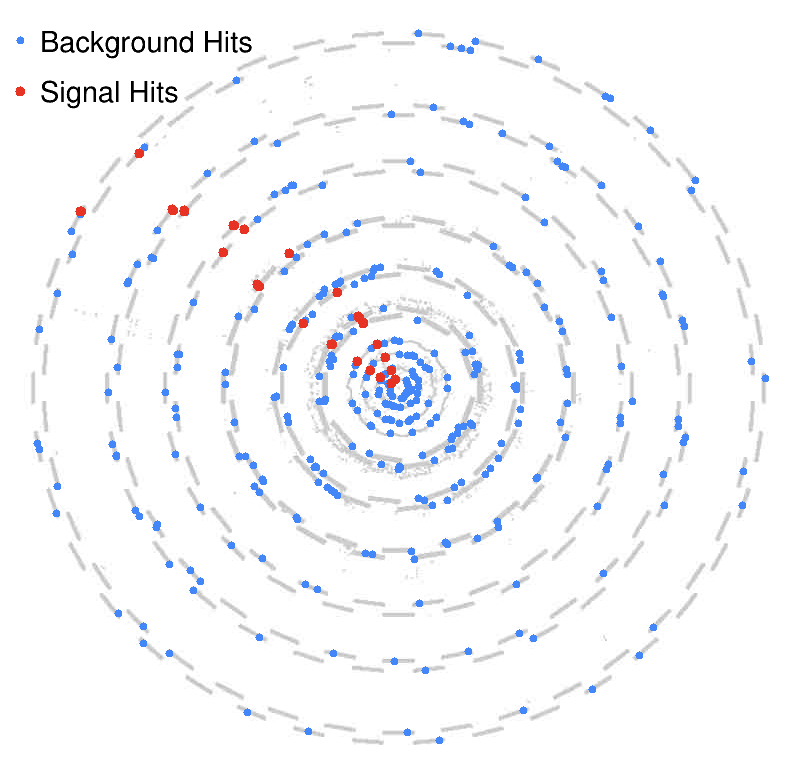}
        \caption{\emph{Example of an event containing two particles with an angular separation of $\Delta\Phi = 300\ \text{mrad}$.}}
        \label{fig:overlapping tracks}
    \end{subfigure}
    \caption{\emph{Study of the network's performance as a function of the angular difference between tracks. As $\Delta \Phi$ decreases, the network's ability to correctly classify the different tracks diminishes, indicating a loss of specialization in the learned representations. The error associated to each point is of order $O(10^{-3}).$}}
    \label{fig:combined_figure}
\end{figure}

\begin{definition}[Misclassification Rate]
    The misclassification rate is the ratio of misclassified events ({\it i.e.}, events when a neuron fires and that do not contain a track of the class it was assigned to) to the total number of events in which the neuron fires in the dataset:
    \begin{equation*}
        \alpha = \frac{FP}{TP + FP}
    \end{equation*}
    where:
    \begin{itemize}
        \item $TP$ (True Positives) are correctly classified particle track events.
        \item $FP$ (False Positives) are events mistakenly classified.
    \end{itemize}
\end{definition}
Figure~\ref{fig:neuron_efficiency_vs_angle} examines the detection efficiency of each neuron versus the angular difference between tracks. For all particle combinations, we observed a significant loss in efficiency as azimuthal angular separation between particle pairs decreases. 
This tendency was confirmed by a concurrent rise in the misclassification rate, represented in Fig.~\ref{fig:misclassification_rate_vs_angle}.
Figure~\ref{fig:angulare_difference_study} is intended as a summary of those observations, displaying the average behavior for efficiency and misclassification rates over all particle types. 
Those results can be ascribed to the refractory period, governed by the dynamics of the reset potential, and the inhibition mechanism that prevents the simultaneous firing of neurons. 
In principle, this effect represents the main limitation of the network in multi-track events. We believe that it can be mitigated by carefully tuning the parameters that regulate the reset potential after neuron activation, as well as by adjusting the strength and duration of inhibition. Furthermore, we believe that an encoding of the so-far neglected transversal dimension (corresponding to particles of different rapidity generating hits at different $z$ positions in the detector) could greatly improve the capability of the SNN to distinguish particles with similar azimuthal angles. These aspects will be further investigated in future studies.

\FloatBarrier

\section{Conclusions} 

In this work we have considered a simplified model of the CMS phase-2 silicon tracker, to study how a neuromorphic computing readout and processing of the information may allow the unsupervised identification of particle trajectories at particle colliders, in the presence of significant noise in the detector.

Our proposed approach introduces several novel concepts, including time-encoding of ionization hits and neuromorphic time-series processing, which could significantly reduce power consumption in online tracking systems. With appropriate hardware implementation, this method has the potential to enable ultra-fast identification of trigger primitives, making it a promising candidate for real-time applications in future high-energy physics experiments.

While in our model we limited ourselves to a 2D geometry and we mainly focused on the issue of single track reconstruction, we identified no conceptual hindrances to the scalability of this model to three spatial dimensions, which would strongly reduce backgrounds, nor to the tracking of large numbers of particles in the same event. One challenging aspect of the produced model is on the other hand the need for tuning of its hyperparameters, in order for the system to achieve the best performance. In our study we explored two different techniques to identify the best operating point of those parameters, through an evolutionary technique and the separate learning of input signal delays to the neurons. Since the amount of training data necessary to learn patterns by the system is quite limited (few tens of thousands of events, corresponding to $O(1 \mu$s) of data taking at the LHC), hyperparameters tuning does not appear as a potential showstopper in the way to a real hardware implementation of a neuromorphic-computing-based triggering system for tracking applications in future colliders.

\bibliographystyle{unsrt}
\bibliography{references.bib}
\clearpage
\section{Supplementary Material and Parameters}
\begin{table}[ht]
    \centering
    \setlength{\extrarowheight}{5pt}
    \begin{tabularx}{\textwidth}{|>{\hspace{2mm}}l<{\hspace{2mm}}|>{\centering\arraybackslash}p{2cm}|X|}
        \hline
        \textbf{Parameter}        & \textbf{Value}                 & \textbf{Description} \\
        \hline
        $K_\mu$       & 0.167             & Temporal Scaling Factor for Inhibitory Post-Synaptic Potential (IPSP) dynamics. \\
        \hline
        K                       & 2.27 V            & Scaling constant for Excitatory Post-Synaptic Potential (EPSP). \\
        \hline
        $K_1$                      & 3.45 V            & Scaling constant for the Reset Potential after neuron activation. \\
        \hline
        $K_2$                      & 5.00 V                          & Scaling constant for the Reset Potential after neuron activation. \\
        \hline
        $d_{\text{max}}$                & $2.5\times 10^{-9}$ s          & Maximum allowable synaptic delay. \\
        \hline
        $\Delta$                & $5\times 10^{-10}$ s          & Defines the initial spread of the synaptic delays. \\
        \hline
        $T_0$               & 1.58 V & Membrane potential threshold for neuron activation in Layer 0. \\
        \hline
        $T_1$ & 0.733 V              & Membrane potential threshold for neuron activation in Layer 1. \\
        \hline
        $\alpha$                & 1.31              & Strength coefficient for Inhibitory Post-Synaptic Potential (IPSP) dynamics. \\
        \hline
        $d_{-}$                 & $1.98 \times 10^{-13}$ s      & Learning rate for synaptic delay depression. \\
        \hline
        $d_{+}$                 & $2.24 \times 10^{-13}$ s      & Learning rate for synaptic delay potentiation. \\
        \hline
        $\tau_m$                & $1.24 \times 10^{-10}$ s      & Membrane potential decay time constant. \\
        \hline
        $\tau_s$                & $3.46 \times 10^{-11}$ s      & Synaptic potential time constant for EPSP dynamics. \\
        \hline
        $\tau_{d_-}$             & $1.31 \times 10^{-9}$ s         & Time constant for synaptic delay depression. \\
        \hline
        $\tau_{d_-}^{'}$            & $2.90 \times 10^{-10}$ s        & Auxiliary time constant for synaptic delay depression. \\
        \hline
        $\tau_{d_+}$             & $2.70 \times 10^{-9}$ s         & Time constant for synaptic delay potentiation. \\
        \hline
        $\tau_{d_+}^{'}$ $\tau_{d_+}^{'}$            & $6.15 \times 10^{-10}$ s        & Auxiliary time constant for synaptic delay potentiation. \\
        \hline
        $t_{max}$               & $6.14 \times 10^{-11}$ s      & Peak time for Excitatory Post-Synaptic Potential (EPSP). \\
        \hline
    \end{tabularx}
    \vspace{0.3cm}
    \caption{\emph{Final Hyperparameters of the Spiking Neural Network, optimized for robust particle trajectory classification under high-noise conditions using a Genetic Algorithm.}}
    \label{tab:hyperparameters}
\end{table}


\end{document}